\documentclass[apj]{emulateapj}

\shorttitle{Redshift Evolution of AGN bias}
\shortauthors{Allevato et al.}

\begin{document}

\slugcomment{Accepted for publication in The Astrophysical Journal}

\title{The XMM-Newton Wide field survey in the COSMOS field:\\
redshift evolution of AGN bias and subdominant role of mergers in triggering moderate luminosity AGN at redshift
up to 2.2}

\author{V. Allevato\altaffilmark{1}, 
A. Finoguenov\altaffilmark{2,5}, 
N. Cappelluti\altaffilmark{8,5}, 
T. Miyaji\altaffilmark{6,7}, 
G. Hasinger\altaffilmark{1}, 
M. Salvato\altaffilmark{1,3,4}, 
M. Brusa\altaffilmark{2}, 
R. Gilli\altaffilmark{8}, 
G. Zamorani\altaffilmark{8}, 
F. Shankar\altaffilmark{14}, 
J. B. James\altaffilmark{10,11}, 
H. J. McCracken\altaffilmark{9}, 
A. Bongiorno\altaffilmark{2}, 
A. Merloni\altaffilmark{3,2}, 
J. A. Peacock\altaffilmark{10}, 
J. Silverman\altaffilmark{12} and 
A. Comastri\altaffilmark{8}}

\altaffiltext{1}{Max-Planck-Institut f\"{u}r Plasmaphysik, Boltzmannstrasse 2, D-85748 Garching, Germany}
\altaffiltext{2}{Max-Planck-Institute f\"{u}r Extraterrestrische Physik, Giessenbachstrasse 1, D-85748 Garching, Germany}
\altaffiltext{3}{Excellent Cluster Universe, Boltzmannstrasse 2, D-85748, Garching, Germany}
\altaffiltext{4}{California Institute of Technology, 1201 East California Boulevard, Pasadena, 91125, CA}
\altaffiltext{5}{University of Maryland, Baltimore County, 1000 Hilltop Circle, Baltimore, MD 21250, USA}
\altaffiltext{6}{Instituto de Astronomia, Universidad Nacional Autonoma de Mexico, Ensenada, Mexico (mailing adress: PO Box 439027, San Ysidro, CA, 92143-9024, USA)}
\altaffiltext{7}{Center for Astrophysics and Space Sciences, University of California at San Diego, Code 0424, 9500 Gilman Drive, La Jolla, CA 92093, USA}
\altaffiltext{8}{INAF-Osservatorio Astronomico di Bologna, Via Ranzani 1, 40127 Bologna, Italy}
\altaffiltext{9}{Observatoire de Paris, LERMA, 61 Avenue de l'Obervatoire, 75014 Paris, France}
\altaffiltext{10}{Astronomy Department, University of California, Berkeley, 601 Campbell Hall, Berkeley CA, 94720-7450, USA}
\altaffiltext{11}{Dark Cosmology Centre, University of Copenhagen, Juliane Maries Vej 30, 2100 Copenhagen, Denmark}
\altaffiltext{12}{Institute for the Physics and Mathematics of the Universe, The University of Tokyo, 5-1-5 Kashiwanoha, Kashiwa-shi, Chiba 277-8583, Japan}
\altaffiltext{14}{Max-Planck-Institute f\"{u}r Astrophysik, Karl-Scwarzschild-Str., D-85748 Garching, Germany}

\begin{abstract}

  We present a study of the redshift evolution of the projected correlation function of 
  593 X-ray selected AGN  with $I_{AB}<23$ and spectroscopic redshifts $z<4$,
  extracted from the 0.5-2 keV X-ray mosaic of the 2.13$deg^2$ XMM-COSMOS survey. 
  We introduce a method to estimate the average bias of
  the AGN sample and the mass of AGN hosting halos, solving the sample variance
  using the halo model and taking into account the growth of the structure over time.
  We find evidence of a redshift evolution of the bias factor for the total population
  of XMM-COSMOS AGN from $\overline{b}(\overline{z}=0.92)=2.30 \pm 0.11$ 
  to $\overline{b}(\overline{z}=1.94)=4.37 \pm 0.27$ with an average
  mass of the hosting DM halos $logM_{0}[h^{-1} M_{\odot}] \sim 13.12 \pm 0.12$
  that remains constant at all $z < 2$.  \\
  Splitting our sample into broad optical lines AGN
  (BL), AGN without broad  optical lines (NL) and X-ray unobscured and obscured AGN, we observe 
  an increase of the bias with redshift in the range $\overline{z}=0.7-2.25$ and 
  $\overline{z}=0.6-1.5$ which corresponds to a constant halo mass 
  $logM_{0}[h^{-1} M_{\odot}] \sim 13.28 \pm 0.07$ and $logM_{0}[h^{-1} M_{\odot}] \sim 13.00 \pm 0.06$
  for BL /X-ray unobscured AGN and NL/X-ray obscured AGN, respectively. \\
 The theoretical models which assume a quasar phase triggered by
 major mergers can not reproduce the high
 bias factors and DM halo masses found for X-ray selected BL AGN with 
 $L_{BOL}\sim 2 \times 10^{45} erg$ s$^{-1}$.
 Our work extends up to $z \sim 2.2$  the $z \lesssim 1$ statement that, for moderate luminosity X-ray selected BL AGN, 
 the contribution from major mergers is outnumbered by other processes, possibly secular such as tidal disruptions or disk instabilities.
\end{abstract}


\section{Introduction}
\label{sec:intro}

Investigating the clustering properties of active galactic nuclei (AGN)
is important to put tight constraints on 
how the AGN are triggered and fueled, to 
identify the properties of the AGN host galaxies, 
and to understand how galaxies and AGN co-evolve.
In addition, in the framework of the cold dark matter (CDM) structure formation
scenario, clustering properties or the bias of AGN, may be related to the
typical mass of dark matter (DM) halos in which they reside \citep{Mo96, She99, 
She01, Tin05} and allow various types of AGN to be placed in a cosmological context.

Recently, several studies have been made, employing spectroscopic redshifts
to measure the three dimensional correlation function of X-ray AGN.  The
majority of the X-ray surveys agree with a picture where X-ray AGN are
typically hosted in DM halos with mass of the order of $12.5 <
logM_{DM}[h^{-1}M_{\odot}] < 13.5$, at low ($z<0.4$) and high ($z\sim 1$)
redshift \citep{Gil05, Yan06, Gil09, Hic09, Coi09, Kru10, Cap10}.
This implies that X-ray AGN more likely reside in massive DM halos
and preferentially inhabit dense environment typical of galaxy groups. 

There have been attempts to detect X-ray luminosity dependence of the clustering. 
At $z \sim 1$, neither \citet{Gil09} nor \citet{Coi09} found significant 
dependence of the clustering amplitudes on the optical luminosity,
X-ray luminosity or hardness ratio, partially due to the larger statistical errors.
Recent works by \citet{Kru10} and \citet{Cap10} found, however, 
that high X-ray luminosity AGN cluster more strongly than low X-ray luminosity ones 
at $2 \sigma $ level for $z \sim 0.3$ and $z \sim 0$, respectively.
\begin{figure*}
\plottwo{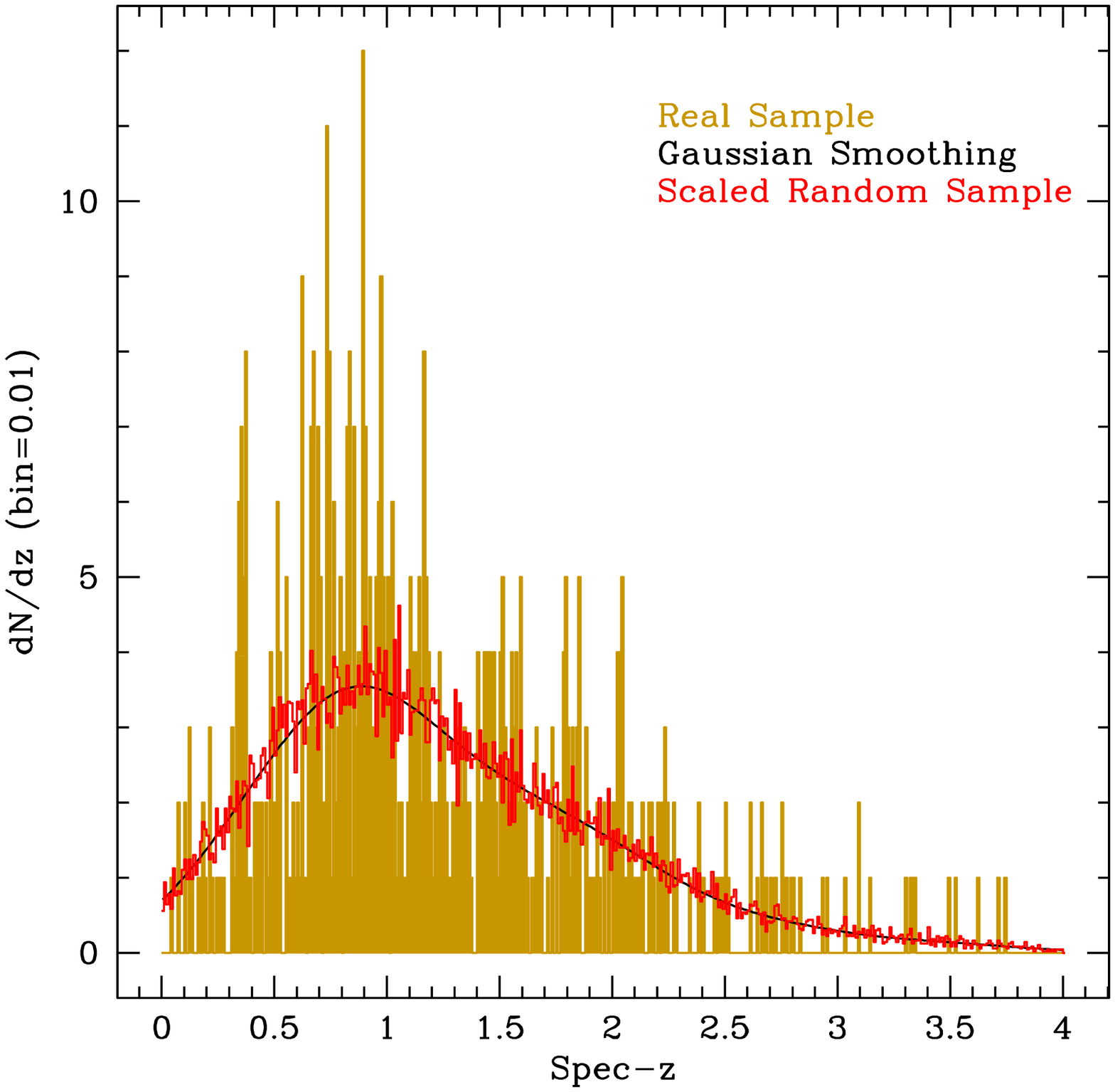}{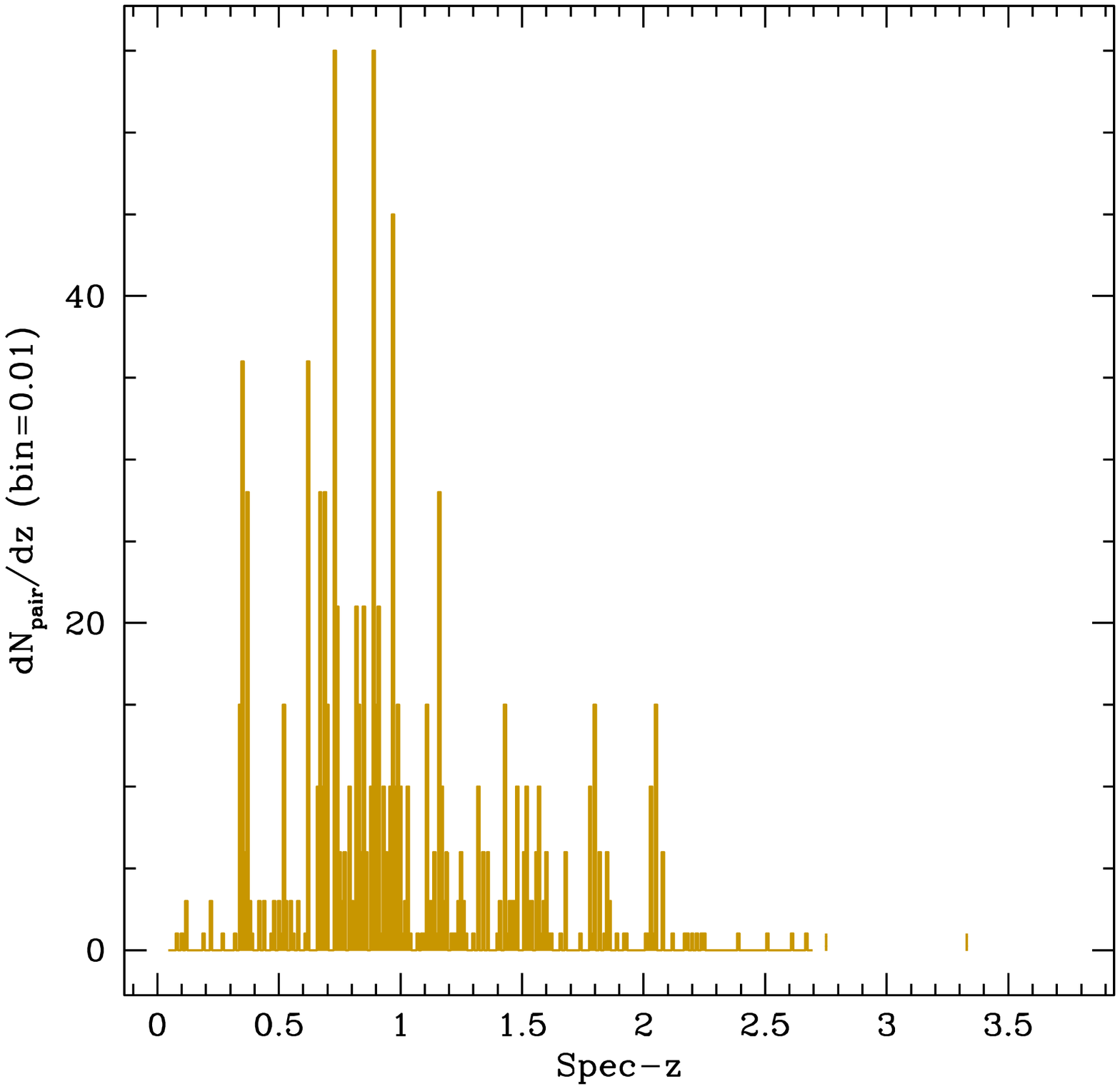}
\caption{\footnotesize \emph{Left panel}: Redshift distribution of 593 AGN
  (gold filled histogram) in bins of $\Delta z=0.01$, with median
  $\overline{z}=1.22$. The solid black curve is the Gaussian smoothing of
  the AGN redshift distribution with $\sigma_z=0.3$, used to
  generate the random sample (red empty histogram). \emph{Right panel}:
  distribution of AGN pairs in redshift bins $\Delta z=0.01$.}
\label{fig:historan}
\end{figure*} 

Until recently, the clustering of AGN has been studied mainly in optical, 
particularly in large area surveys such as 2dF \citep[2QZ,][]{Cro05, Por06}
and Sloan Digital Sky Survey \citep[SDSS,][]{Li06, Shen09, Ros09}.
\citet{Cro05} analysed the clustering of 2QZ QSO as a function of redshift
finding a strong evolution of QSO bias, with $b_{Q}(z=0.53)=1.13 \pm 0.18$
at low redshift and $b_{Q}(z=2.48)=4.24 \pm 0.53$ at high redshift,
as also observed in \citet{Por06}.
The evidence of an evolution over time of the bias factor for SDSS quasars has been
found in \citet{Shen09}, with bias values ranging from $b_{Q}(z=0.50)=1.32 \pm 0.17$ to
$b_{Q}(z=3.17)=7.76 \pm 1.44$.
The results from these surveys have also shown that the bias evolution of optically selected 
quasars is consistent with an approximately constant mass at all redshifts of the hosting
DM halo in the range $logM_{DM} \sim 12.5-13[h^{-1}M_{\odot}]$.\\
Besides models of major mergers between gas-rich galaxies appear
to naturally produce the bias of quasars as
a function of $L$ and $z$ \citep{Hop08, Shen09a, Sha09, Sha10, Sha10rev, Bon09},
supporting the observations that bright quasars host galaxies present a preference
for merging systems. 
It is still to be verified if the results from optical surveys can be extended 
to the whole AGN population and in particular to the X-ray selected AGN. 

In this paper, we concentrate on the study of the bias evolution with redshift
using different X-ray AGN samples and we focus on the estimation of the bias factor 
and the hosting halo mass using a new method which properly account for the 
sample variance and the strong evolution of the bias with the time.\\
The paper is organized as follows. In section \S\ref{sec:cat} we describe the XMM-COSMOS AGN sample
and the AGN subsets used to estimate the correlation function. 
In \S\ref{sec:rand} we describe the random
catalog generated to reproduce the properties of the data sample and
the method to measure two-point statistic is explained in \S\ref{sec:meth}. 
The results of the AGN auto-correlation based on the standard method of
the power-law fitting of the signal
and using the two-halo term are given in \S\ref{sec:results}. In \S\ref{sec:newmeth}
we present our own method to estimate the AGN bias factor and the DM halos masses in
which AGN reside, solving the sample variance and the bias evolution with redshift 
and in \S \ref{sec:Mea} the results. In \S\ref{sec:bevol} we
present the redshift evolution of the bias factor and the corresponding 
DM halo masses for the different AGN subsets.
We discuss the results in the context of previous studies
in \S\ref{sec:disc} and we conclude in \S\ref{sec:conc}. Throughout the
paper, all distances are measured in comoving coordinates and are given in
units of Mpc $h^{-1}$, where $h=H_0/100$ km/s. We use a $\Lambda$CDM cosmology with
$\Omega_M=0.3$, $\Omega_\Lambda=0.7$, $\Omega_b=0.045$, $\sigma_8=0.8$.
The symbol $log$ signifies a base-10 logarithm.

\begin{figure*}
\plottwo{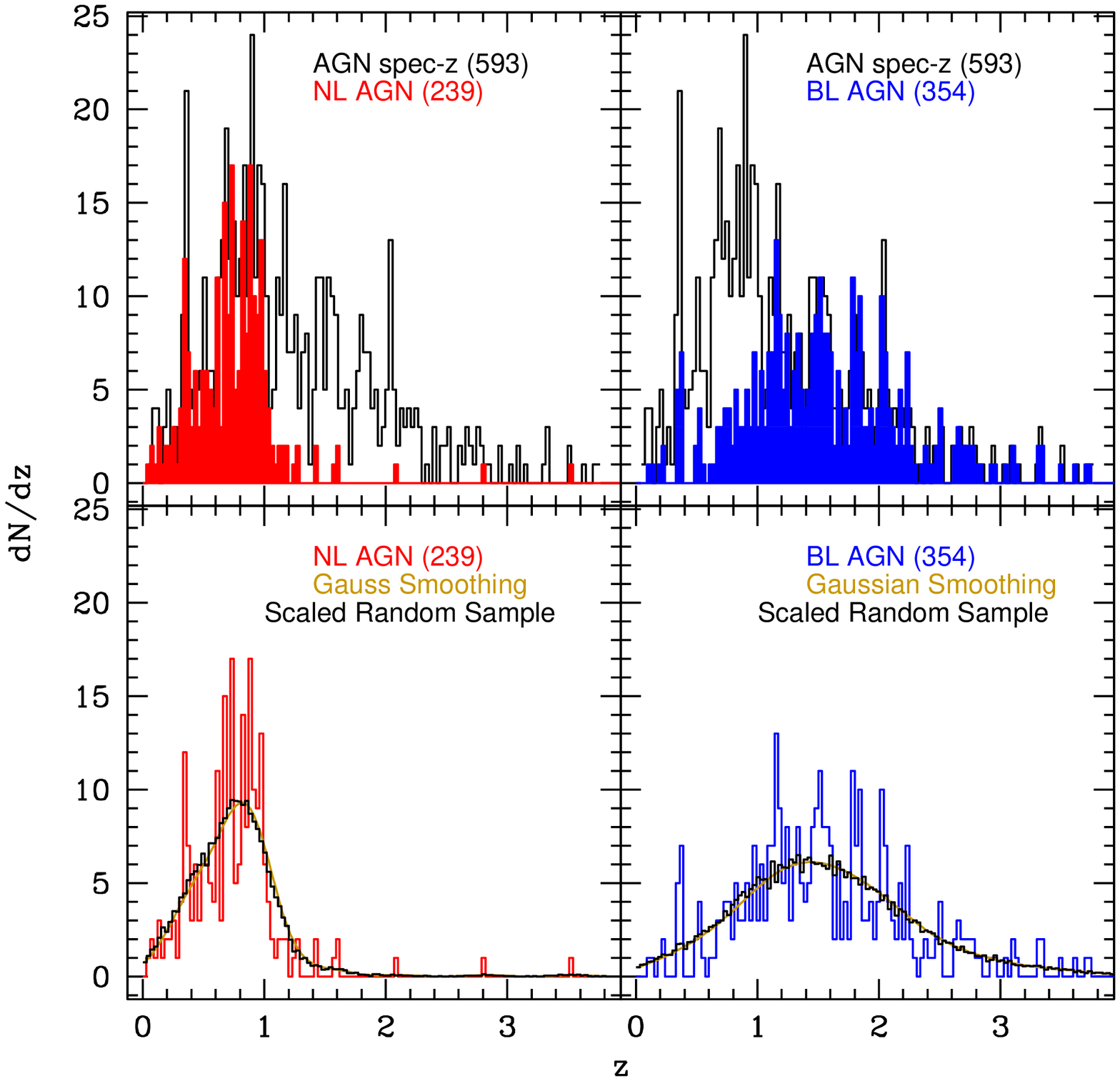}{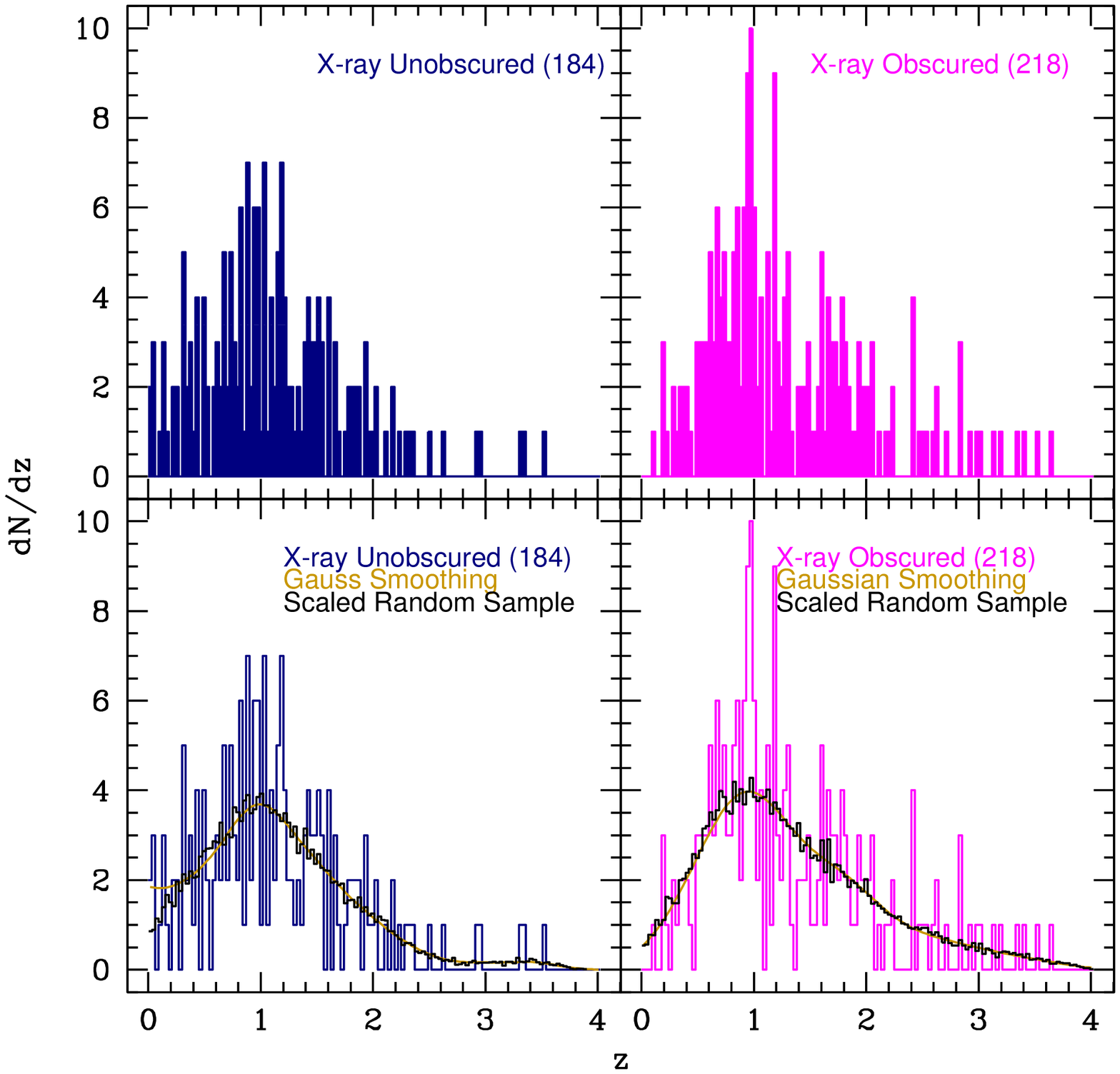}
\caption{\footnotesize \emph{Left Panel}: Redshift distribution of XMM-COSMOS AGN (open
  histogram) selected in the soft band, compared with the redshift distribution of BL AGN (blue
  histogram, \emph{upper right quadrant}) and NL AGN, (red, \emph{upper left
    quadrant}). Lower quadrants show the redshift distribution of the random
  catalogs (open black histograms) for both the AGN sub-samples, obtained
  using a Gaussian smoothing (gold lines) of the redshift distribution of
  the real samples. \emph{Right Panel}: Redshift distribution of unobscured (dark blue histogram) and obscured (magenta histogram)
  AGN selected in the hard band according with the column density (\emph{upper quadrants}). Lower quadrants show 
  the redshift distribution of the random catalogs (open black histograms) for both the AGN sub-samples, obtained
  using a Gaussian smoothing (gold lines) of the redshift distribution of the real samples.} 
\label{fig:blnldistr}
\end{figure*}

\section{AGN Catalog}
\label{sec:cat}

The \emph{Cosmic Evolution Survey} (COSMOS) is a multiwavelength
observational project over $1.4 \times 1.4$ $deg^2$ of equatorial field
centred at $(RA,DEC)_{J2000}$ = (150.1083, 2.210), aimed to study AGN,
galaxies, large scale structure of the Universe and their co-evolution. The
survey uses multi wavelength imaging from X-ray to radio
bands, including HST \citep{Sco07}, SUBARU \citep{Tan07}, Spitzer \citep{San07}
and GALEX \citep{Zam07}. The central 0.9 $deg^2$ of the COSMOS field has been observed in X-ray with
\emph{Chandra} for a total of 1.8 Ms \citep{Elv09}.
In addiction spectroscopic campaigns have been carried out
with VIMOS/VLT and extensive spectroscopic follow-up have been granted
with the IMACS/Magellan, MMT and DEIMOS/KeckII projects.\\
XMM-\emph{Newton} surveyed $2.13$ $deg^2$ of
the sky in the COSMOS field in the 0.5-10 keV energy band for a total of
$\sim$ 1.55 Ms \citep{Has07,Cap07, Cap09} providing an unprecedently 
large sample of point-like X-ray sources (1822).\\
The XMM-COSMOS catalog has been cross-correlated 
with the optical multiband catalog \citep{Cap07}, the K-band catalog 
\citep{McC10}, the IRAC catalog \citep{San07, Ilb09} and the MIPS catalog \citep{LeF09}.
\citet{Bru10} presented the XMM-COSMOS multiwavelength catalog
of 1797 X-ray sources with optical/near infrared identification,
multiwavelength properties and redshift information \citep[from][]{Lil07, Lil09, Tru07, 
Ade06, Pre06, Sal09}.

In this paper we focus on the clustering analysis of 1465 XMM-COSMOS AGN
detected in the energy band 0.5-2 keV, for which we have a
spectroscopic completeness of $ \sim 53 \%$ (780/1465).
From this sample of 780 objects we selected 593 sources
with $I_{AB}<23$ (this magnitude cut increases
the spectroscopic completeness to about 65\%) and redshift $z<4$. 
The redshift distribution of the
AGN sample (Fig. \ref{fig:historan} \emph{left panel}) shows prominent
peaks at various redshifts, $z \sim 0.12$, $z \sim 0.36$, $z \sim 0.73$, $z\sim 0.95$,
$z\sim 1.2$, $z\sim 2.1$. In particular, the structure at $z \sim 0.36$ was
also observed at other wavelengths in COSMOS \citep{Lil07} and already
discussed \citep{Gil09}. The median redshift of the sample is $<z>=1.22$. \\
The sources have been classified in \citet{Bru10} in broad optical line AGN (BL AGN, 354),
non-broad optical line AGN (NL AGN, 239) using a combination of
X-ray and optical criteria, motivated by the fact that both obscured and
unobscured AGN can be misclassified in spectroscopic studies, given that the
host galaxy light may over shine the nuclear emission. Fig.  \ref{fig:blnldistr}
shows the redshift distribution of BL AGN with 
$<z>=1.55$ and NL AGN with $<z>=0.74$.\\
We also studied the clustering properties of X-ray unobscured and obscured AGN
derived on the basis of the observed X-ray hardness ratio
and corrected to take into account the redshifts effects.
In particular we used the hard X-ray band (2-10 keV) (which allows us to sample 
the obscured AGN population)
to select a subset of 184 X-ray unobscured sources (X-unobs hereafter) 
with $log N_H < 22$ $cm^{-2}$ and 218 X-ray obscured (X-obs hereafter) sources with $log N_H \geq 22$ $cm^{-2}$,
The median redshift of the two sub-samples are $<z>=1.12$ and $<z>=1.30$, 
respectively (see fig. \ref{fig:blnldistr}, \emph{right panel}).
The 47\% (40\%) of BL (NL) AGN have been also observed 
in the hard band and classified as X-unobs (X-obs) AGN.

\section{Random Catalog}
\label{sec:rand}

The measurements of two-point correlation function 
requires the construction of a random catalog with the same
selection criteria and observational effects as the data, to serve as an
unclustered distribution to which to compare. XMM-Newton observations
have varying sensitivity over the COSMOS field. In order to create an AGN random
sample, which takes the inhomogeneity of the sensitivity over the field into
account, each simulated source is placed at random position in the sky, with
flux randomly extracted from the catalog of real sources fluxes (we verified
that such flux selection produces the same results as if extracting the
simulated sources from a reference input logN-logS). The simulated source is
kept in the random sample if its flux is above the sensitivity map value at
that position \citep{Miy07, Cap09}. Placing
these sources at random position in the XMM-COSMOS field has 
the advantage of not removing the
contribution to the signal due to angular clustering. 
On the other hand, this procedure does not take into account
possible positional biases related to the optical follow-up program.
\citet{Gil09}, who instead decided to extract the coordinates
of the random sources from the coordinate ensemble of the read sample,
showed that there is a difference of only 15\% in the correlation
lengths measured with the two procedures. 
 
The corresponding redshift for a random object is assigned based on the
smoothed redshift distribution of the AGN sample. As in \citet{Gil09} 
we assumed a Gaussian smoothing length $\sigma_z = 0.3$. This is
a good compromise between scales that are either too small, thus affected by local density variations 
or too large and thus oversmooth the distribution (our results do not change significantly using
$\sigma_z=0.2-0.4$). Fig. \ref{fig:historan} (\emph{left panel}) shows the
redshift distribution of 593 XMM-COSMOS AGN and the scaled random sample ($\sim 41000$
random sources) which follows the red solid curve obtained by Gaussian
smoothing.

\section{Two-point Statistics}
\label{sec:meth}

A commonly used technique for measuring the spatial clustering of a class of
objects is the two-point correlation function $\xi(r)$, which measures the excess
probability $dP$ above a random distribution of finding an object in a
volume element $dV$ at a distance $r$ from another randomly chosen object
\citep{pee80}:
\begin{equation}
dP = n[1 + \xi(r)] dV
\end{equation}
where $n$ is the mean number density of objects. 
\begin{figure}[!ht]
\epsscale{1.}
\plotone{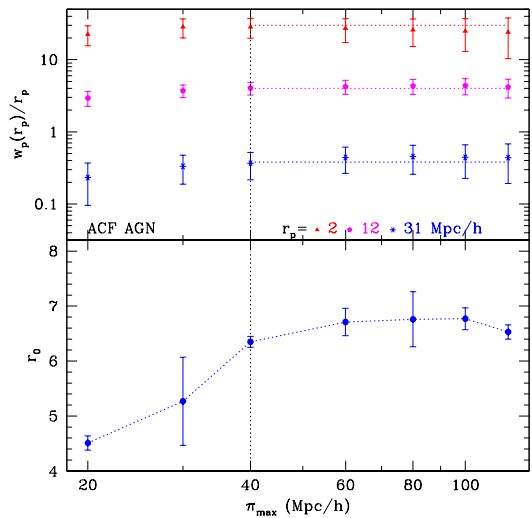}
\caption{\footnotesize Projected AGN correlation function $w_p(r_p)$
  computed at different $r_p$ scale (see label) as function of the integral
  radius $\pi_{max}$. Horizontal lines show that the ACF saturates for
  $\pi_{max} >$40 Mpc/h, which is also the minimum $\pi_{max}$ at which
  $w_p(r_p)$ converges and returns the smaller error on the best-fit
  correlation parameter $r_0$, with $\gamma$ fixed to 1.8.} 
\label{fig:pimaxroprovaa}
\end{figure}
In particular, the auto-correlation function (ACF) measures the excess probability of
finding two objects from the same sample in a given volume element.
With a redshift survey, we cannot directly
measure $\xi(r)$ in physical space, because peculiar motions of galaxies
distort the line-of-sight distances inferred from redshift. To separate the
effects of redshift distortions, the spatial correlation function is
measured in two dimensions $r_p$ and $\pi$, where $r_p$ and $\pi$ are the
projected comoving separations between the considered objects in the
directions perpendicular and parallel, respectively, to the mean
line-of-sight between the two sources.  Following \citet{Dav83},
$r_1$ and $r_2$ are the redshift positions of a pair of objects, $s$ is the
redshift-space separation $(r_1 - r_2)$, and $l = \frac{1}{2}(r_1+r_2)$ is
the mean distance to the pair. The separations between the two considered
objects across $r_p$ and $\pi$ are defined as:
\begin{eqnarray}
\pi & = & \frac{\textbf{s}\cdot \textbf{l}} {|\textbf{l}|}\\
r_p & = & \sqrt{(\textbf{s} \cdot \textbf{s} - \pi^2)}
\end{eqnarray}
 Redshift space distortions only affect the correlation function along the line of sight, 
 so we estimate the so-called projected correlation function $w_p(r_p)$ \citep{Dav83}:
\begin{eqnarray}\label{eq:integral}
w_p(r_p) = 2 \int_0^{\pi_{max}} \xi(r_p,\pi) d\pi 
 \end{eqnarray}
 where $ \xi(r_p,\pi)$ is the two-point correlation function in term of
 $r_p$ and $\pi$, measured using the \citet[LS]{Lan93} estimator:
\begin{equation}\label{eq:LZ}
\xi = \frac{1}{RR'} [DD'-2DR'+RR']
\end{equation}
DD', DR' and RR' are the normalized data-data, data-random and random-random number of pairs defined by:
\begin{eqnarray}
DD' = \frac{DD(r_p,\pi)}{n_d(n_d - 1)}\\
DR' = \frac{DR(r_p,\pi)}{n_dn_r}\\
RR' = \frac{RR(r_p,\pi)}{n_r(n_r-1)}
\end{eqnarray}
where $DD$, $DR$ and $RR$ are the number of data-data, data-random and
random-random pairs at separation $r_p \pm \Delta r_p$ and $\pi \pm \Delta
\pi$ and $n_d$, $n_r$ are the total number of sources in the data and random
sample, respectively. Fig. \ref{fig:historan} (\emph{right panel}) shows the number of pairs
in redshift bins $\Delta z=0.01$ for the AGN sample. 

\begin{figure*}
\plottwo{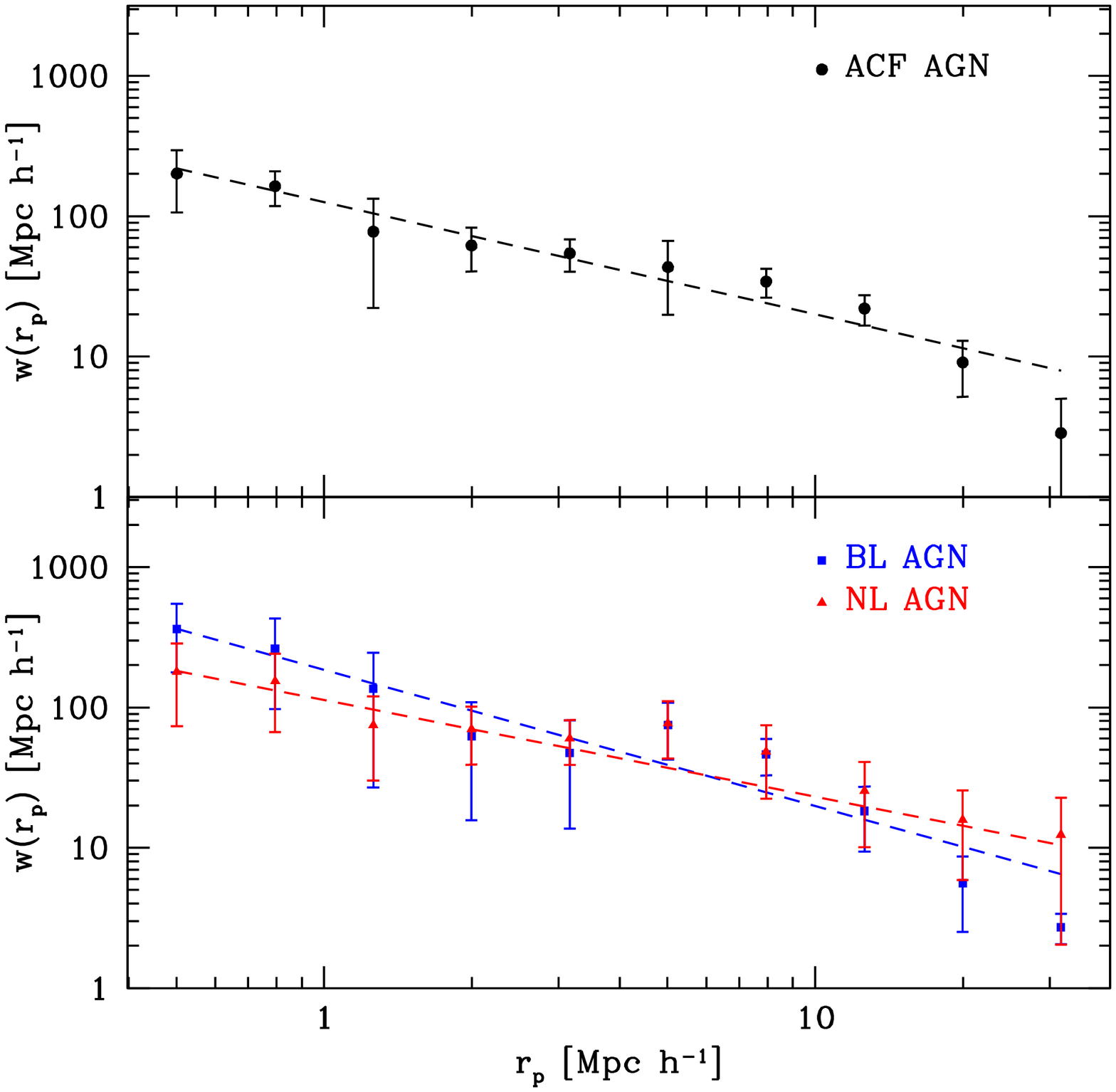}{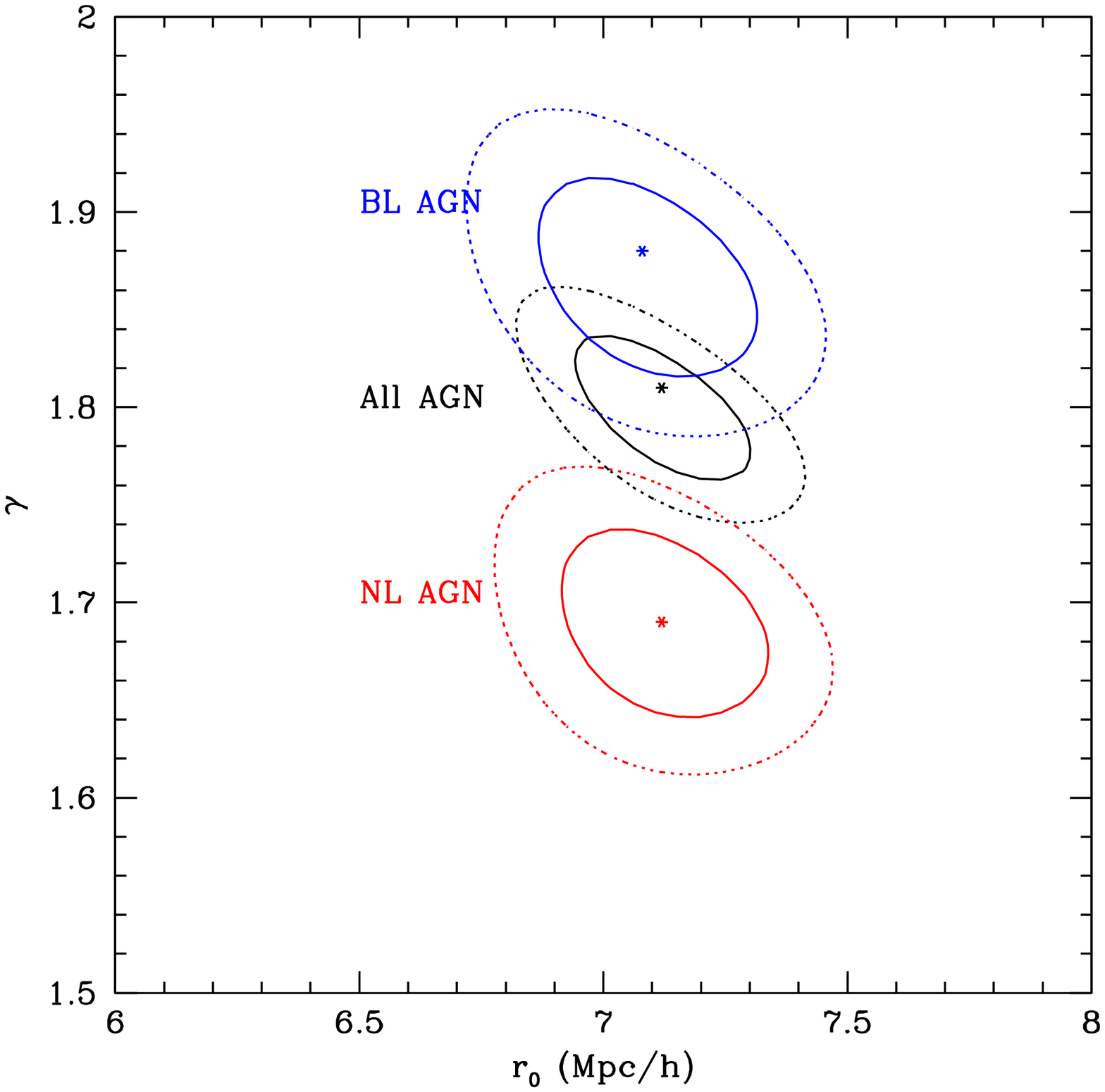}
\caption{\footnotesize \emph{Left panel}: Projected AGN ACF 
  (black circles) compared to the auto-correlation of BL AGN (blue squares)
  and NL AGN (red triangles). The data points are fitted with a power-law model
  using the $\chi^2$ minimization technique; the errors are computed with a
  bootstrap resampling method. \emph{Right panel}: The confidence contours
  of the power-law best-fit parameters $r_0$ and $\gamma$, for the whole AGN
  sample (black), for the BL AGN (blue) and NL AGN (red) sub-samples. 
  The contours mark the 68.3\% and 95.4\% confidence levels (respectively corresponding to
  $\Delta \chi^2 = 2.3$ and 6.17) are plotted as continuous and dotted lines.}
\label{fig:acfagn1}
\end{figure*}
The LS estimator has been used to measure correlations in a number of surveys,
for example, SDSS \citep{Zeh05, Li06}, DEEP2 \citep{Coi07, Coi08}, AGES 
\citep{Hic09}, COSMOS \citep{Gil09}. If $\pi_{max}=\infty$, then we 
average over all line-of-sight peculiar velocities, and $w_p(r_p)$ 
can be directly related to $\xi(r)$ for a power-law parameterization, by:
\begin{equation}\label{eq:model}
w_p(r_p) = r_p \left( \frac{r_0}{r_p} \right)   ^{\gamma} \frac{\Gamma(1/2) \Gamma[(\gamma - 1)/2] }{\Gamma(\gamma/2)} 
\end{equation}
In practice, we truncate the integral at a finite $\pi_{max}$ value, to maximize the correlation signal. 
One should avoid values of $\pi_{max}$ too large since they would add noise
 to the estimate of $w_p(r_p)$; if instead, $\pi_{max}$ 
 is too small one would not recover all the signal.
To determine the appropriate $\pi_{max}$ values for the XMM-COSMOS AGN correlation
function,  we estimated $w_p(r_p)$ for different values of $\pi_{max}$ in
the range 20-120 Mpc $h^{-1}$. Besides, we determined the correlation length $r_0$
for this set of $\pi_{max}$  values, by fitting $w_p(r_p)$ with a fixed
$\gamma$=1.8 over $r_p$ in the range 0.5-40 Mpc $h^{-1}$. In
Fig. \ref{fig:pimaxroprovaa} we show the increase of the projected
AGN auto-correlation $w_p(r_p)$ as a function of the integration radius
$\pi_{max}$. The $w_p(r_p)$ values appear to converge for $\pi_{max}>40$
Mpc $h^{-1}$. Therefore we adopt $\pi_{max}$= 40 Mpc $h^{-1}$ in the following analysis, which
is the minimum $\pi_{max}$ at which the correlation function
converges. Such $\pi_{max}$ selection returns the smallest error on the
best-fit correlation parameter $r_0$.

\section{Projected Auto-correlation Function}
\label{sec:results}

\subsection{Standard Approach}
\label{sec:powlaw}

To estimate the AGN auto-correlation function $\xi(r_p,\pi)$ using the LS
formula (Eq. \ref{eq:LZ}), we created a grid with $r_p$ and $\pi$ in the
range 0.1-100 Mpc $h^{-1}$, in logarithmic bins $\Delta log (r_p,\pi) = 0.2$ and we
projected $\xi(r_p,\pi)$ on $r_p$ using Eq. \ref{eq:integral}.\\
In literature, several methods are adopted for error estimates in two-point
statistics and no one has been proved to be the most precise. It is known
that Poisson estimators generally underestimate the variance because they do
not account for the fact that the points are not statistically independent, i.e. the same
objects appear in more than one pair.  In this work we computed the errors
on $w_p(r_p)$ with a bootstrap resampling technique \citep{Coi09, Hic09,
Kru10, Cap10}.

The standard approach used to evaluate the power of the clustering
signal is to fit $w_p(r_p)$ with a power-law model \citep{Coi09, Hic09,
Gil09, Kru10, Cap10} of the form given in Eq.
\ref{eq:model}, using a $\chi^2$ minimization technique, with $\gamma$ and
$r_0$ as free parameters.
Fig. \ref{fig:acfagn1} (\emph{left panel, upper quadrant}) shows the
projected AGN ACF, evaluated in the projected separation
range $r_p$= 0.5-40 Mpc $h^{-1}$. The best-fit correlation length and 
slope and the corresponding $1\sigma$ errors, are found to be $r_0 = 7.12^{+0.28}_{-0.18}$ Mpc $h^{-1}$
and $\gamma=1.81^{+0.04}_{-0.03}$. \\
We estimated the projected correlation function of BL and NL AGN in the range $r_p=0.5-40$ Mpc $h^{-1}$,
as shown in Fig.\ref{fig:acfagn1} (\emph{left panel, lower quadrant}). 
For BL AGN we found a correlation length $r_0=7.08^{+0.30}_{-0.28}$
Mpc $h^{-1}$ and $\gamma=1.88^{+0.04}_{-0.06}$, while for NL AGN we
measured $r_0=7.12^{+0.22}_{-0.20}$ Mpc $h^{-1}$ and a flatter slope
$\gamma=1.69^{+0.05}_{-0.05}$. 
Fig. \ref{fig:acfagn1} (\emph{right panel}) shows the power-law best-fit
parameters for the different AGN samples with the 1$\sigma$ and 2$\sigma$
confidence intervals for a two parameter fit, which correspond to
$\chi^2=\chi_{min}^2+2.3$ and $\chi^2=\chi_{min}^2+6.17$.  

We can estimate
the AGN bias factor using the power-law best fit parameters:
\begin{equation}\label{eq:PLb}
b_{PL}=\sigma_{8,AGN}(z)/\sigma_{DM}(z)
\end{equation}
where $\sigma_{8,AGN}(z)$ is
rms fluctuations of the density distribution over the sphere with a comoving
radius of 8 Mpc $h^{-1}$, $\sigma_{DM}(z)$ is the DM correlation function 
evaluated at 8 Mpc $h^{-1}$, normalized to a value of $\sigma_{DM}(z=0)=0.8$. For a
power-law correlation function this value can be calculated by \citep{pee80}:
\begin{equation}\label{eq:bias}
(\sigma_{8,AGN})^{2} = J_2(\gamma)(\frac{r_0}{8 Mpc/h})^{\gamma}
\end{equation}
where $J_2(\gamma)=72/[(3-\gamma)(4-\gamma)(6-\gamma)2^{\gamma}]$.  
As the linear regime of the structure formation is verified only 
at large scales, the best-fit parameters $r_0$ and $\gamma$ are estimated 
fitting the projected correlation function on $r_p=1-40$
Mpc $h^{-1}$.  The $1\sigma$ uncertainty of $\sigma_{8,AGN}$ is computed from the $r_0$ vs.
$\gamma$ confidence contour of the two-parameter fit corresponding to
$\chi^{2}=\chi^{2}_{min}+2.3$. 

\subsection{Two-halo Term}
\label{sec:2halo}

\begin{deluxetable}{lccccccc}
\tabletypesize{\scriptsize}
\tablewidth{0pt}
\tablecaption{Bias Factors and hosting DM halo masses \label{tbl-1}}
\tablehead{
\colhead{(1)} &
\colhead{(2)} &
\colhead{(3)} &
\colhead{(4)} &
\colhead{(5)} \\
\colhead{$AGN$} &
\colhead{$<z>$\tablenotemark{a}} &
\colhead{$b_{PL}$} &
\colhead{$b_{2-h}$} & 
\colhead{$log\overline{M}_{DM}$\tablenotemark{b}}\\
\colhead{$Sample$} &
\colhead{} &
\colhead{Eq. \ref{eq:PLb}} &
\colhead{Eq. \ref{eq:b}} & \colhead{$h^{-1}M_{\odot}$} } 
\startdata
\multicolumn{8}{c}{}\\
Total (593)& 1.22 & $2.80^{+0.22}_{-0.90}$ & $2.98 \pm 0.13$ & $13.23 \pm 0.06$  \\
BL (354)& 1.55 & $3.11^{+0.30}_{-1.22}$ & $3.43 \pm 0.17$ & $13.14 \pm 0.07$  \\
NL (239)& 0.74 & $2.78^{+0.45}_{-1.07}$ & $2.70 \pm 0.22$ & $13.54 \pm 0.10$ \\
X-unobs (184)& 1.12 & $2.98^{+0.34}_{-0.37}$ & $3.01 \pm 0.21$ & $13.33 \pm 0.08$  \\
X-obs (218)& 1.30 & $1.66^{+0.31}_{-0.32}$ & $1.80 \pm 0.15$ & $12.30 \pm 0.15$  \\
\multicolumn{8}{c}{Subsample at $z<1$}\\
BL (70)& 0.57 & $2.18^{+0.95}_{-1.02}$ & $2.32 \pm 0.26$ & $13.50 \pm 0.11$  \\
NL (137)& 0.53 & $1.68^{+0.45}_{-0.57}$ & $1.40 \pm 0.15$ & $12.65 \pm 0.18$  \\
\enddata
\tablenotetext{a}{Median redshift of the sample.}
\tablenotetext{b}{Typical DM halo masses based on \citet{She01} and \citet{van02}.}
\end{deluxetable}

\begin{figure}
\plotone{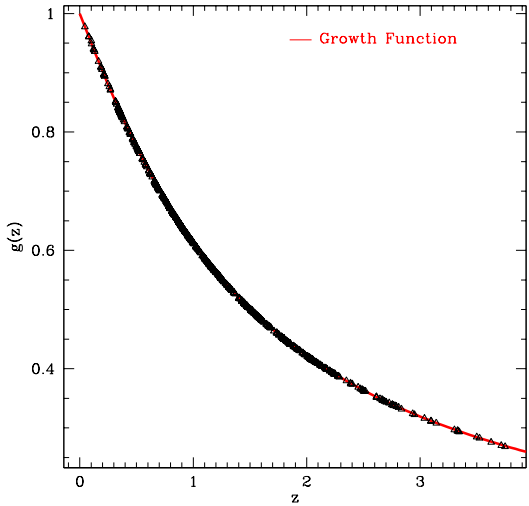}
\caption{\footnotesize Factor $g$ as defined in Eq. \ref{eq:gz}, estimated at the redshift of each AGN (black triangles). The data points are fitted by the function $D_1(z)/D_1(z=0)$, where $D_1(z)$ is the growth function (see Eq. (10) in 
Eisenstein \& Hu 1999 and references therein). The bias of each AGN is weighted by this factor according to the redshift $z$ of the source.}
\label{fig:gz}
\end{figure}
In the halo model approach, the two-point
correlation function  of AGN is the sum of two contributions: the first
term (1-halo term) is due to the correlation between objects in the
same halo and the second term (2-halo term) arises because of the
correlation between two distinct halos: 
\begin{equation}
w_{AGN}(r_p) = w_{AGN}^{1-h}(r_p)+w_{AGN}^{2-h}(r_p)
\end{equation}
As the 2-halo term dominates at large scales, we can
consider this term to be in the regime of linear density fluctuations. 
In the linear regime, AGN are biased tracers of the DM
distribution and the AGN bias factor defines the relation
between the two-halo term of DM and AGN.
\begin{equation}
w^{2-h}_{AGN}(r_p)=b^{2}_{AGN}w^{2-h}_{DM}(r_p)
\end{equation}
We first estimated the DM 2-halo term at the median redshift of the sample, using:
\begin{equation}\label{eq:2-halo}
\xi^{2-h}_{DM}(r)=\frac{1}{2\pi^2}\int P^{2-h}(k)k^2 \left[ \frac{sin(kr)}{kr} \right]  dk
\end{equation}
where $P^{2-h}(k)$ is the Fourier Transform of the linear power spectrum,
assuming a power spectrum shape parameter $\Gamma = 0.2$ and
$h=0.7$.
Following \citet{Ham02}, we estimated 
$\xi^{2-h}_{DM}(r)$ and then the DM projected correlation $w_{DM}^{2-h}(r_p)$ using:
\begin{equation}
w_{DM}^{2-h}(r_p)=2 \int_{r_p}^{\infty} \frac{\xi^{2-h}_{DM}(r)rdr}{\sqrt{r^2-r_p^2}}
\end{equation}
\begin{figure}
\plotone{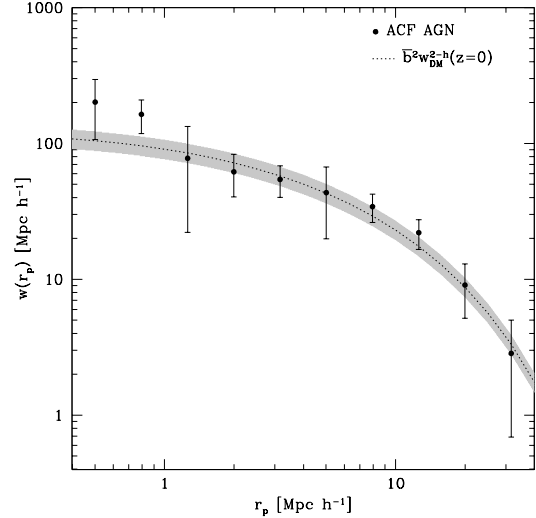}
\caption{\footnotesize Projected AGN ACF (black circles) compared to $\overline{b}^2w_{DM}^{2-h}(r_p,z=0)$ (dotted line), where the weighed bias $\overline{b}$ is defined in Eq. \ref{eq:weigb}. The shaded region shows the projected DM 2-halo term scaled by $(\overline{b} \pm \delta \overline{b})^2$.}
\label{fig:AGNhalo}
\end{figure}
 
\begin{figure*}
\plottwo{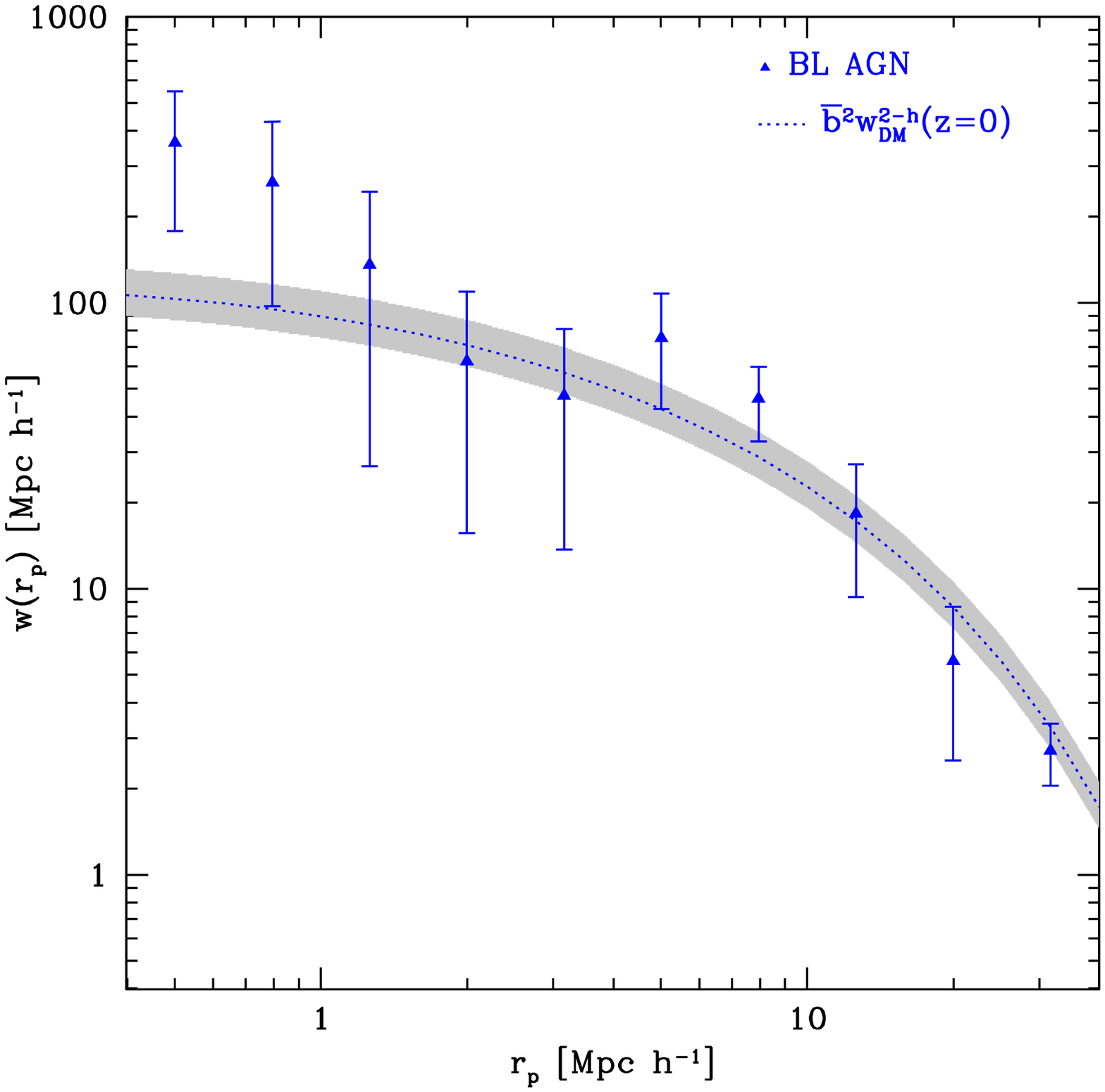}{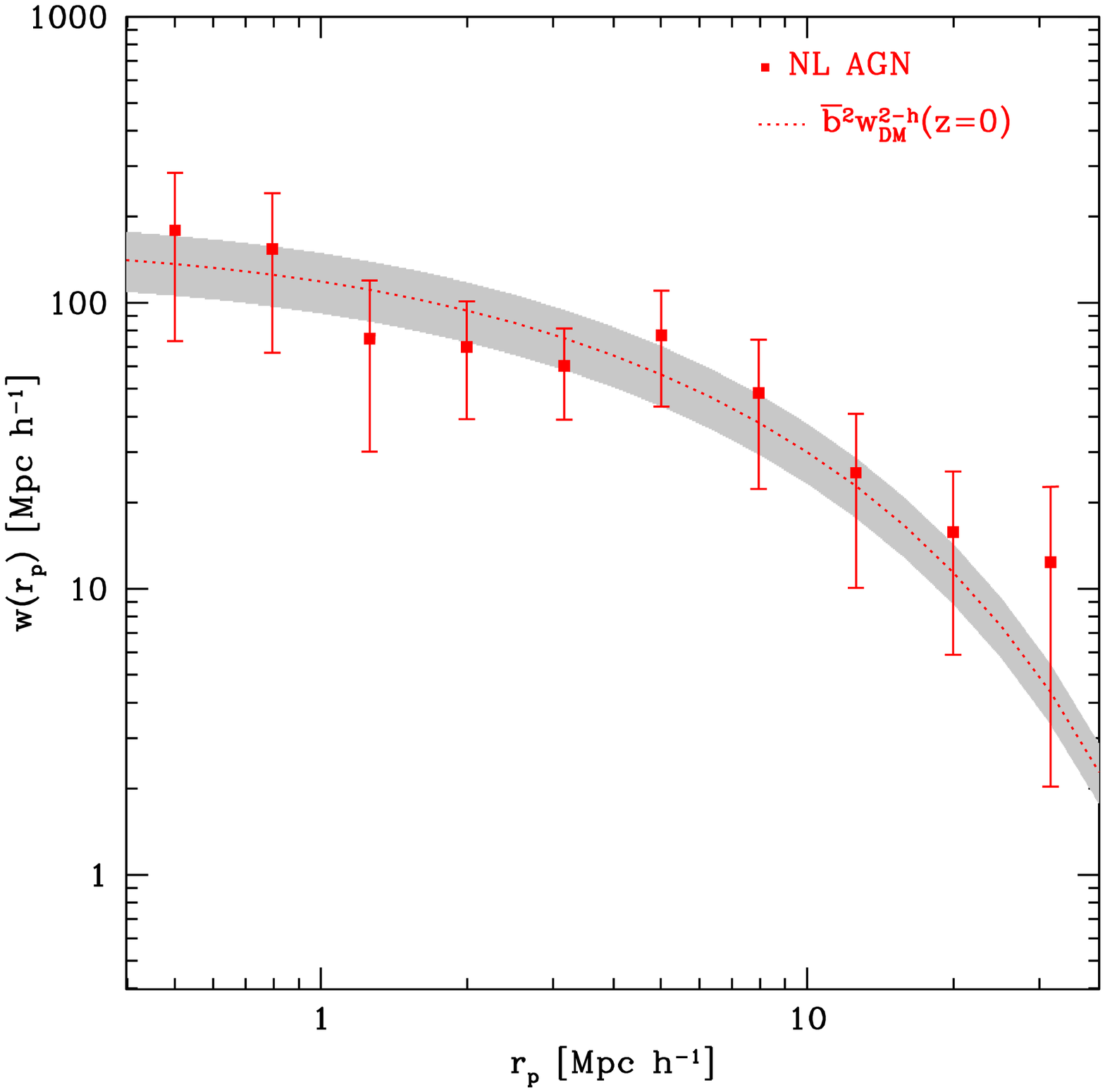}
\caption{\footnotesize Projected ACF of BL AGN (blue triangles, \emph{left panel}) and NL AGN (red squares, \emph{right panel}), compared to $\overline{b}^2w_{DM}^{2-h}(r_p,z=0)$ (dotted line), where the weighed bias $\overline{b}$ is defined in Eq. \ref{eq:weigb}. The shaded region shows the projected DM 2-halo term scaled by $(\overline{b} \pm \delta \overline{b})^2$.}
\label{fig:blnlhalo}
\end{figure*} 

\begin{figure*}
\plottwo{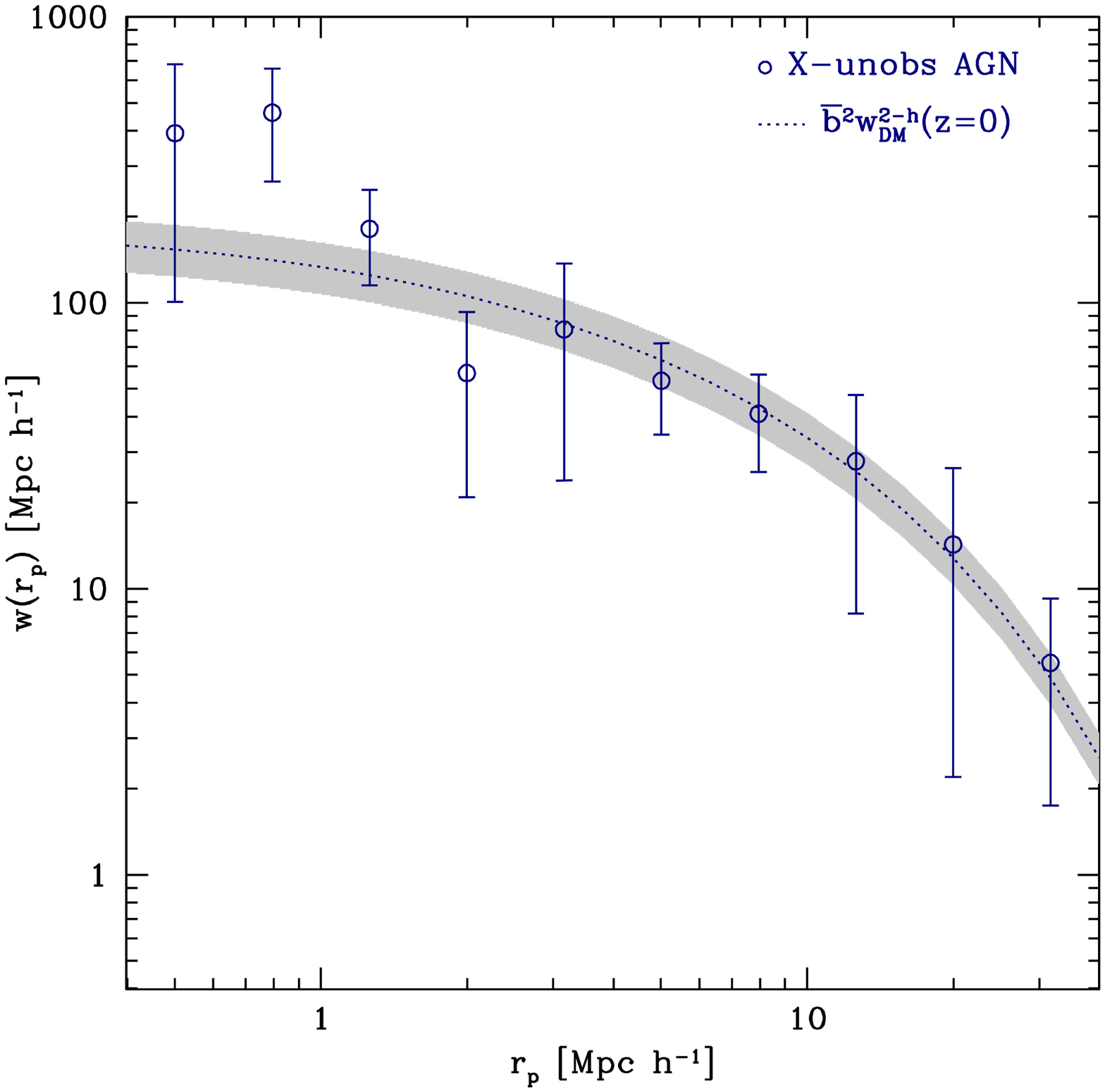}{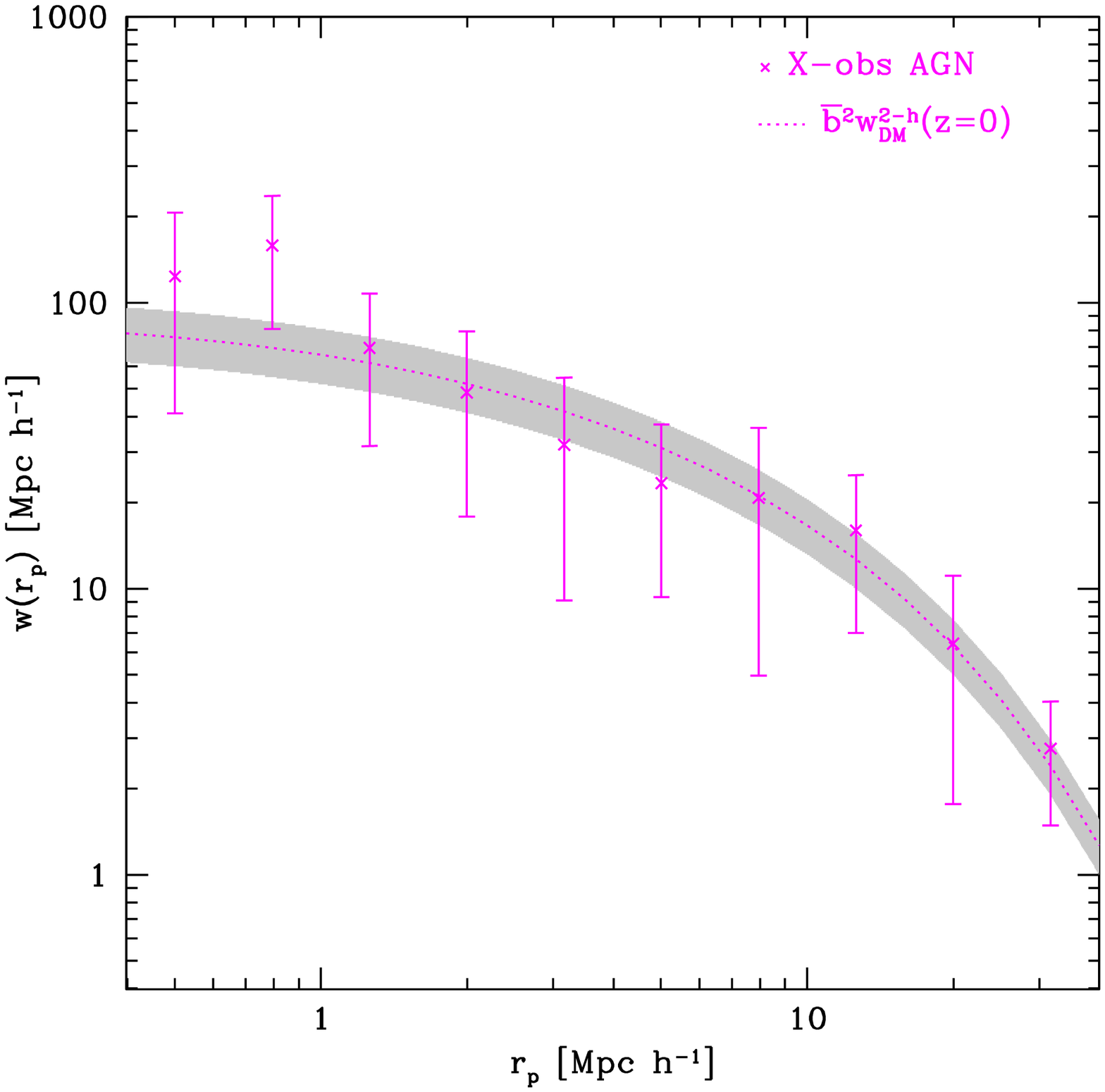}
\caption{\footnotesize Projected ACF of X-unobs AGN (darkblue open circles, \emph{left panel}) and X-obs AGN (magenta diagonal crosses, \emph{right panel}), compared to $\overline{b}^2w_{DM}^{2-h}(r_p,z=0)$ (dotted line), where the weighed bias $\overline{b}$ is defined in Eq. \ref{eq:weigb}. The shaded region shows the projected DM 2-halo term scaled by $(\overline{b} \pm \delta \overline{b})^2$.}
\label{fig:ty12halo}
\end{figure*} 
Using this term, we can estimate the AGN bias simply dividing
the projected AGN correlation function at large scale ($r_p>1$ Mpc $h^{-1}$)
by the DM 2-halo term:
\begin{equation}\label{eq:b}
b^2_{AGN}=(w_{AGN}{2-h}(r_p)/w_{DM}^{2-h}(r_p))^{1/2}
\end{equation}
and then averaging 
over the scales $r_p=1-40$ Mpc $h^{-1}$.
Table \ref{tbl-1}, column 4 shows the AGN bias factors 
using this method, compared with the
ones based on the power-law fits of the ACF (column 3) for the different AGN subsets.
The two sets of bias values from the different approaches are consistent within $1\sigma$,
but the errors on $b_{PL}$ are bigger consistently with the fact that the
AGN ACF is not well described by a power-law.

\section{Solving for Sample Variance using HOD}
\label{sec:newmeth} 

\begin{deluxetable}{lcccc}
\tabletypesize{\scriptsize}
\tablewidth{0pt}
\tablecaption{Weighted Bias factors and hosting DM halo masses \label{tbl-2}}
\tablehead{
\colhead{(1)} &
\colhead{(2)} &
\colhead{(3)} &
\colhead{(4)} &
\colhead{(5)} \\
\colhead{$AGN$} &
\colhead{$\overline{b}$} &
\colhead{$\overline{z}$} &
\colhead{$logM_{0}$} &
\colhead{$b_{S01}$\tablenotemark{a}} \\
\colhead{$Sample$} &
\colhead{Eq. \ref{eq:weigb}} &
\colhead{Eq. \ref{eq:weigz}} &
\colhead{$h^{-1}M_{\odot}$} & \colhead{} }
\startdata
\multicolumn{5}{c}{}\\
Total (593)& $1.91\pm 0.13$ & 1.21 & $13.10 \pm 0.06$& $2.71 \pm 0.14$  \\
BL (354)& $1.74\pm 0.17$ & 1.53 & $13.24 \pm 0.06$ & $3.68 \pm 0.27$ \\
NL (239) & $1.80\pm 0.22$ & 0.82 & $13.01 \pm 0.08$ & $2.00 \pm 0.12$ \\
X-unobs (184)& $1.95\pm 0.21$ & 1.16 & $13.30 \pm 0.10$ & $3.01 \pm 0.26$ \\
X-obs (218)& $1.37\pm 0.15$ & 1.02 & $12.97 \pm 0.08$ & $2.23 \pm 0.13$ \\
\multicolumn{5}{c}{Subsample at $z<1$}\\
BL (70)& $1.62\pm 0.26$ &  0.63 & $13.27 \pm 0.10$ & $1.95 \pm 0.17$ \\
NL (137)& $1.56\pm 0.15$ & 0.60 & $12.97\pm 0.07$ & $1.62 \pm 0.15$ \\
\enddata
\tablenotetext{a}{Bias estimated from $M_{0}$ using \citet{She01}.}
\end{deluxetable}

The standard approaches used in previous works on
clustering of X-ray AGN \citep{Mul04, Yan06, Gil05, Coi09, Hic09,
Kru10, Cap10} to estimate the bias factors from the projected AGN ACF 
are based on the power-law fit parameters (method 1).
This method assumes that the projected correlation function is well
fitted by a power-law and the bias factors are derived from
the best fit parameters $r_{0}$ and $\gamma$ of the clustering signal at large scale.\\
Most of the authors \citep{Hic09, Kru10, Cap10}
used  an analytical expression \citep[as the one described in][]{She99,
She01, Tin05} to assign a characteristic DM halo mass to the hosting halos. 
The incongruity of this approach is that the bias used is the
average bias of a given sample at a given redshift. 
However, the average bias is sensitive to 
the entirety of the mass distribution so that
distributions with different average masses, can give rise to the same
average bias value. \\
In the halo model approach the large scale amplitude
signal is due to the correlation between objects in distinct halos
and the bias parameter defines the 
relation between the large scale clustering amplitude 
of the AGN ACF and the DM 2-halo term (method 2).  

In literature, the common model used for the AGN HOD is a three parameter model
including a step function for the HOD of central AGN and a truncated power-law satellite
HOD \citep[introduced by][for galaxies]{Zeh05}. Here we assumed
that all the AGN reside in central galaxies. This assumption is supported
by \citet{Sta10}. They found that X-ray AGN are predominantly 
located in the central galaxies of the host DM halos and tend to avoid 
satellite galaxies, fixing the limit to the fraction of AGN
in non-central galaxies to be less than 10\%. The same fraction of
satellites galaxies hosting AGN is suggested in \citet{Shen09a}.
\citet{Sha10} modelled the measurements of quasar 
clustering derived in the SDSS \citep{Shen09}
and they verified that the predicted bias factors and the correlation functions
are not altered including subhalos as quasar hosts. 
A further consideration is that there is in practice no 
distinction between central and satellite AGN in the
2-halo term that we used to estimate the AGN bias factor. 

We assumed a simple parametric form of the AGN halo occupation $N_{A}$,
described by a delta function:
\begin{equation}\label{eq:AGNHOD}
N_A(M_{DM})=f_A\delta(M_{DM}-M_{0})
\end{equation}
where $f_A$ is the AGN duty cycle. It is clear that we are not
considering the full HOD model, but we are assigning to all the 
AGN the same average mass of the hosting halos. 
The motivation is that X-ray AGN mainly reside in massive halos
with a narrow distribution of the hosting halo masses. It's clear
that this assumption is specific to AGN and e.g. is not applicable
to galaxies.

The AGN HOD descrived by $\delta$-function is motivated 
by the results of \citet{Miy11} showing that the AGN HOD rapidly decreases at high halo
masses. In addition \citet{Mar09} and \citet{Sil09} found that AGN 
preferentially reside in galaxy groups rather than in clusters.\\
The $\delta$-function is the simplest possible
assumption in the treatment of the sample variance, 
which is due to the variation in the amplitude of source counts
distribution. It has been shown in \citet{Fal10}, 
that the variation in the density field, which is responsible for the
sample variance, can be replaced by the variation of the halo mass function. 
In terms of halo model, the bias factor as a function of the fluctuations 
$\Delta$ in the density field is expressed by:
\begin{equation}\label{eq:bAGN}
b_{A}(\Delta)= \frac{\int_{M_{h}} N_A(M_{h})b_{h}(M_{h})n_{h}(M_{h},\Delta)dM_{h}}{\int_{M_{h}} N_{A}(M_h)n_{h}(M_{h},\Delta)dM_{h}}
\end{equation}
where $N_A$ is the AGN HOD,
$b_h(M_{DM})$ is the halo bias and $n_{h}(M_{DM},\Delta)$ is the halo mass function,
which depends on the density field.
On the other hand the sample variance does not effect the 
AGN halo occupation.
In Allevato et al. (in prep.) we confirm the assumption of constancy
of the AGN HOD with the density field. \\
When we assume that all AGN reside in DM halos
with the same mass, Eq. \ref{eq:bAGN} becomes simpler:
\begin{equation}
\frac{\int_{M_{h}} \delta(M_{h}-M_0)b_{h}(M_{h})n_{h}(M_{h},\Delta)dM_{h}}{\int_{M_{h}} \delta(M_h-M_0)n_{h}(M_{h},\Delta)dM_{h}}=b(M_0)
\end{equation}
The equation shows that when the AGN HOD is close to
a $\delta$-function, the variations in the density
field only change the AGN number density
and put more weight on AGN bias at 
the redshift of large scale structure (LSS), but do
not change the bias of AGN inside the structure.
Our claim differs from the results presented in \citet{Gil05} and \citep{Gil09}.
They found that excluding sources located within a large-scale structures, 
the correlation length and then the bias factor strongly reduces.
Such bias behaviour can be used to constrain
more complicated shapes of the AGN HOD than a $\delta$-function type distribution.

However, even in the case of a $\delta$-function HOD, we still
need to consider the two effects which are often
omitted in the clustering analysis: the LSS growth and the
evolution of the bias factor with $z$. Ignoring these
effects can by itself lead to a difference in the results reported for
the different AGN samples.
 
The bias factor depends on the redshift as the structures grow over time,
associated with our use of a large redshift interval. 
For the $i^{th}$ source at redshift $z_i$, we considered
the bias factor corresponding to a halo mass $M_{DM}=M_{0}$:
\begin{equation}
b_i=b(M_0,z_i)
\end{equation}
where $b(M_0,z)$ is evaluated using \citet{van02} and \citet{She01}. 
For each AGN at redshift $z$ we estimated 
the factor $g(z)$ defined as the square root of the projected
DM 2-halo term at redshift $z$ normalized to the projected DM 2-halo term
evaluated at $z=0$:
\begin{equation}\label{eq:gz}
g(z)=\sqrt{\frac{w_{DM}(z,r_p)}{w_{DM}(z=0,r_p)}}
\end{equation} 
averaged over the scales $r_p=1-40$ Mpc $h^{-1}$. 
As the amplitude of the projected DM 2-halo term decreases with increasing redshift, 
$g$ is a decreasing function of $z$ (see fig. \ref{fig:gz}), well described by the term
$D_1(z)/D_1(z=0)$, where $D_1(z)$ is the growth function (see eq. (10) in 
\citet{Eis99} and references therein).\\
By accounting for the fact that the linear regime of the structure 
formation is verified only at large scales, 
we estimated the AGN bias considering only the pairs which contribute 
to the AGN clustering signal at $r_p=1-40$ Mpc $h^{-1}$.
We defined the \emph{weighted} bias factor of the sample as:
\begin{equation}\label{eq:weigb}
\overline{b}(M_{0}) = \sqrt{\frac{\sum_{i,j}b_ib_j g_ig_j } {N_{pair}} }
\end{equation}
where $b_ib_j$ is the bias factor of the $i^{th}$ and
$j^{th}$ source in the pair $i-j$, $g_ig_j$ is the $g$
factor of the pair and $N_{pair}$ is the total number of pairs 
in the range $r_p=1-40$ Mpc $h^{-1}$.\\
Similarly, we defined a weighted average redshift of the AGN sample,
weighting the redshift of each pair for the $g$ factor and 
the bias of the pair ($b_ib_j$):
\begin{equation}\label{eq:weigz}
\overline{z}= \frac{\sum_{i,j}b_ib_j g_ig_j z_{pair} } {\sum_{i,j}b_ib_jg_ig_j}
\end{equation} 
where $z_{pair}=(z_i+z_j)/2$.
Following this approach we can find the value of $M_0$ that satisfies:
$$b_1= \overline{b}(M_0) $$
where $b_1$ is the square root of the projected AGN ACF 
normalized to the projected DM 2-halo term at $z=0$:
\begin{equation}
b_1=\sqrt{\frac{w_{AGN}(r_p)}{w_{DM}(z=0,r_p)}}
\end{equation}
averaged over the scale $r_p=1-40$ Mpc $h^{-1}$.

By performing the test in narrow redshift intervals, we can
study the dependency of the halo mass $M_0$ on redshift 
(see \S\ref{sec:bevol}). Moreover with just a single
measurement of the amplitude of the 2-halo term, one cannot
constrain the AGN HOD. Already with several measurements
sampling different density fields, the shape of the HOD can be
linked to the LSS density-dependence of the bias. In addition,
the 1-halo term of the AGN auto-correlation and AGN-groups
cross-correlation can be used to discriminate between different
HOD models, which will be argument of our following work.

\section{Measurements}
\label{sec:Mea}

The weighted bias factors $\overline{b}$ and redshifts $\overline{z}$, and 
the corresponding DM halo masses $M_0$ estimated for the 
different AGN sub-samples using the method described in the
previous section are shown in Table \ref{tbl-2}. \\
Fig. \ref{fig:AGNhalo}, \ref{fig:blnlhalo}  and \ref{fig:ty12halo}
show the ACF of the AGN,
BL/NL AGN and X-unobs/obs AGN samples, 
compared to the term $\overline{b}^2w_{DM}^{2-h}(r_p,z=0)$ (dotted line), 
where the weighed bias $\overline{b}$ is defined in Eq. \ref{eq:weigb}. 
The shaded region shows the projected DM 2-halo term 
scaled by $(\overline{b} \pm \delta \overline{b})^2$.

\begin{figure*}
\plottwo{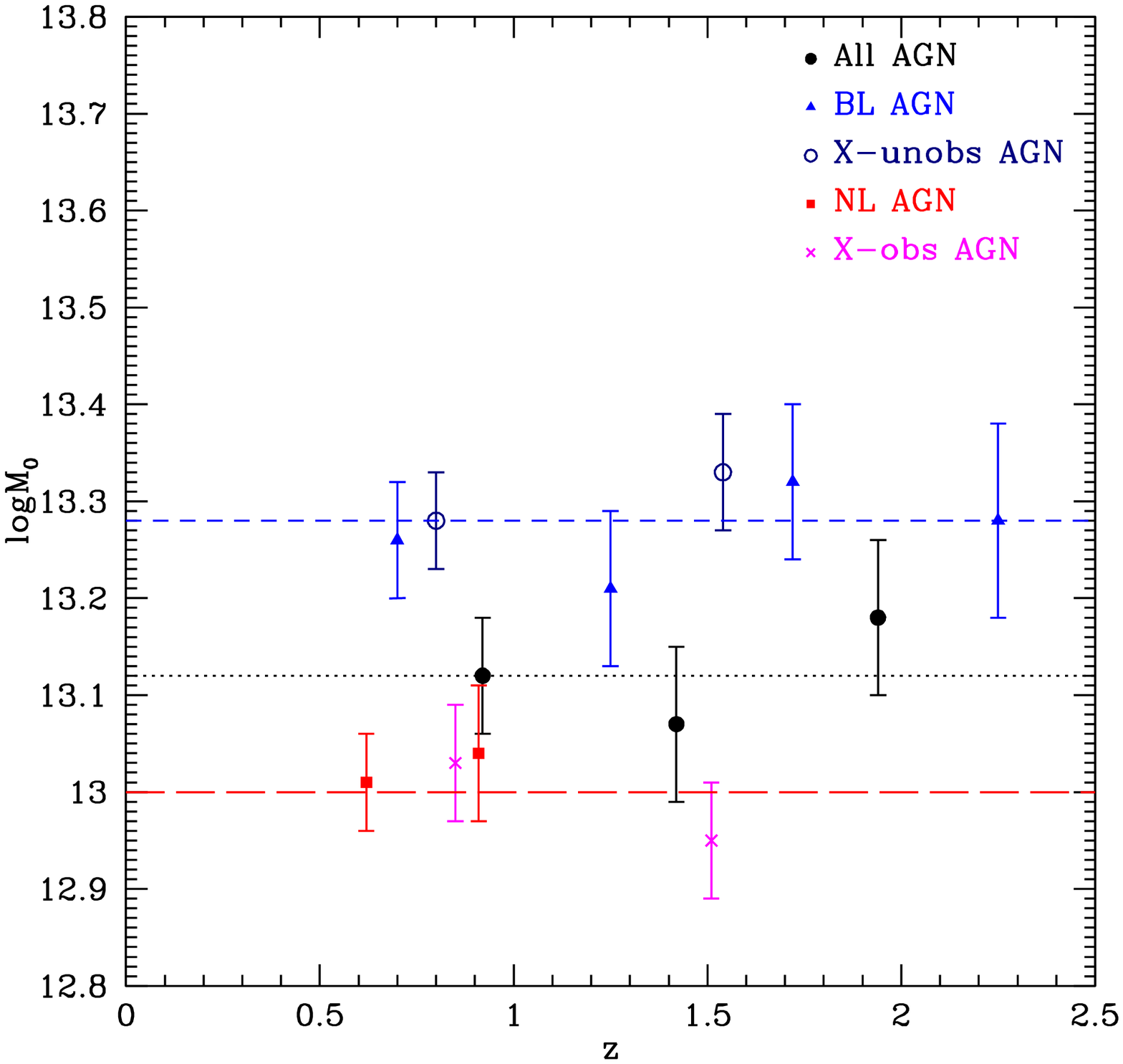}{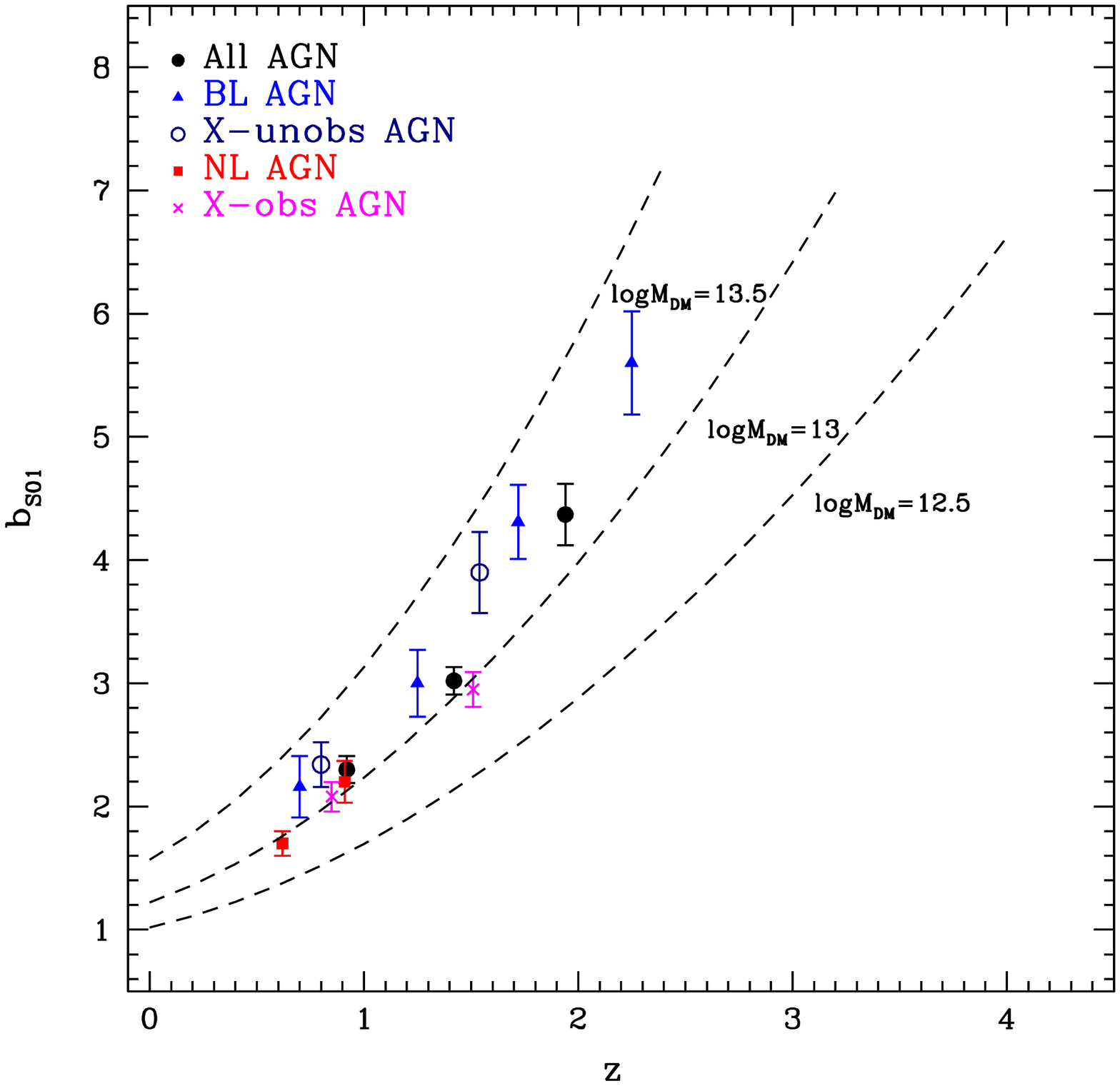}
\caption{\footnotesize \emph{Left Panel}: DM halo mass $M_{0}$ as a function of $z$
for different AGN sub-samples (see legend). The horizontal lines show the mean value of $M_{0}$
for BL/X-unobs AGN (dashed-blue), NL/X-obs AGN (long dashed-red) and for the whole AGN
sample (dotted-black). \emph{Right Panel}: Redshift evolution of the bias parameter $b_{S01}$
of different AGN sub-samples. The dashed lines show the expected
  $b(z)$ of typical DM halo masses $M_{DM}$ based on \citet{She01}.
   The masses are given in $logM_{DM}$ in units of $h^{-1} M_{\odot}$. BL/X-unobs AGN
   present a strong bias evolution with redshift with a constant DM halo mass  
   $logM_{0}=13.28 \pm 0.07 [h^{-1} M_{\odot}]$ up to $\overline{z} \sim 2.4$. NL/X-obs AGN reside in less massive halos
   with $logM_{0}=13.00 \pm 0.06 [h^{-1} M_{\odot}]$, constant at $\overline{z}<1.5$.}
\label{fig:bevol}
\end{figure*}
\begin{deluxetable*}{cccccccc}
\tablewidth{0pt}
\tablecaption{Bias Evolution \label{tbl-3}}
\tablehead{
\colhead{(1)} &
\colhead{(2)} &
\colhead{(3)} &
\colhead{(4)} &
\colhead{(5)} &
\colhead{(6)} &
\colhead{(7)} &
\colhead{(8)} \\
\colhead{$<z>$\tablenotemark{a}} &
\colhead{\textbf{N} } &
\colhead{$b_{2-h}$} &
\colhead{$log\overline{M}_{DM}$\tablenotemark{b}} &
\colhead{$\overline{b}$} &
\colhead{$\overline{z}$} &
\colhead{$logM_{0}$} &
\colhead{$b_{S01}$\tablenotemark{c}} \\
\colhead{} &
\colhead{} &
\colhead{Eq. \ref{eq:b}} &
\colhead{$h^{-1}M_{\odot}$} &
\colhead{} &
\colhead{} &
\colhead{Eq. \ref{eq:b}} & \colhead{$h^{-1}M_{\odot}$}  } 
\startdata
\multicolumn{8}{c}{\em All AGN}\\
0.80 & 190 & $2.70 \pm 0.19$ & $13.48 \pm 0.10$ & $1.80\pm 0.19$ & 0.92 & $13.12 \pm 0.06$ & $2.30 \pm 0.11$ \\
1.30 & 220 & $3.10 \pm 0.18$ & $13.21 \pm 0.10$ & $2.14\pm 0.18$ & 1.42 & $13.07 \pm 0.08$ & $3.02 \pm 0.11$ \\
2.07 & 183 & $5.18 \pm 0.21$ & $13.30 \pm 0.11$ & $2.63\pm 0.21$ & 1.94 & $13.18 \pm 0.08$ & $4.37 \pm 0.27$ \\
\multicolumn{8}{c}{\em BL AGN}\\
0.67 & 70 & $2.62\pm 0.20$ & $13.57 \pm 0.10$ & $1.52\pm 0.20$ & 0.70 & $13.26 \pm 0.06$ & $2.16 \pm 0.25$ \\
1.25 & 108 & $3.06\pm 0.23$ & $13.24 \pm 0.08$ & $2.02\pm 0.23$ & 1.25 & $13.21 \pm 0.08$ & $3.00 \pm 0.27$ \\
1.71 & 92 & $5.37\pm 0.28$ & $13.60 \pm 0.08$ & $3.57\pm 0.28$ & 1.72 & $13.32 \pm 0.08$ & $4.31 \pm 0.30$ \\
2.46 & 85 & $6.82\pm 0.27$ & $13.41 \pm 0.10$ & $4.02\pm 0.27$ & 2.25 & $13.28 \pm 0.10$ & $5.60 \pm 0.42$ \\
\multicolumn{8}{c}{\em X-unobscured AGN}\\
0.65 & 98 & $2.46 \pm 0.17$ & $13.51 \pm 0.11$ & $1.62\pm 0.17$ & 0.80 & $13.28 \pm 0.05$ & $2.34 \pm 0.18$ \\
1.66 & 86 & $4.85 \pm 0.18$ & $13.51 \pm 0.10$ & $2.10\pm 0.18$ & 1.54 & $13.33 \pm 0.06$ & $3.90 \pm 0.33$ \\
\multicolumn{8}{c}{\em NL AGN}\\
0.53 & 137 & $1.40 \pm 0.13$ & $12.65 \pm 0.12$ & $1.59\pm 0.13$ & 0.62 & $13.01 \pm 0.05$ & $1.70 \pm 0.10$ \\
1.02 & 102 & $2.11 \pm 0.19$ & $12.88 \pm 0.15 $ & $1.87\pm 0.19$ & 0.91 & $13.04 \pm 0.07$ & $2.20 \pm 0.17$ \\
\multicolumn{8}{c}{\em X-obscured AGN}\\
0.73 & 106 & $1.80 \pm 0.14$ & $13.01 \pm 0.11 $& $1.51\pm 0.14$ & 0.85 & $13.03 \pm 0.06$ & $2.08 \pm 0.12$ \\
1.84 & 112 & $3.51 \pm 0.16$ & $ 12.94 \pm 0.13 $ & $1.96\pm 0.16$ & 1.51 & $12.95 \pm 0.06$ & $2.95 \pm 0.14$ \\
\enddata
\tablenotetext{a}{Median redshift of the sample.}
\tablenotetext{b}{Typical DM halo masses based on \citet{She01} and \citet{van02}.}
\tablenotetext{c}{Bias estimated from $M_{0}$ using \citet{She01}.}
\end{deluxetable*} 
 
The AGN bias factor indicates
that XMM-COSMOS AGN reside in halos with average mass $logM_{0}=13.01\pm 0.09
[h^{-1}logM_{\odot}]$, characteristic of moderate-size poor groups, a result
consistent with previous works on X-ray selected AGN that indicate that the
typical DM halo mass hosting AGN is in the range $12.5 \lesssim
logM_{DM}\lesssim 13.5[h^{-1}M_{\odot}] $.\\
We found that BL and NL AGN which peak at $\overline{z}=1.53$ and 
$\overline{z}=0.82$, present consistent bias factors
which correspond to DM halo average masses
$logM_{0}=13.24\pm 0.06 [h^{-1} M_{\odot}]$ and
$13.01\pm 0.08 [h^{-1} M_{\odot}]$, respectively.
As described in \citet{Bru10}, only a small fraction of the objects classified as NL
AGN are located at $z>1$, to be compared with 350 in the BL AGN sample. This
is mostly due to the fact that high-redshift NL AGN are optically faint
(typically $I \sim 23-24$) and have not been targeted yet with dedicated
spectroscopic campaigns.  Our results might be affected by the limitations in the
obscured AGN classification, considering that some models on the evolution
of the obscured AGN fraction predict an increase of the fraction with the
redshift \citep{Has08}. In order to avoid the problem 
of different redshift distribution in comparing
BL/NL AGN clustering amplitude, we selected for each sample a subset
(BL AGN with 70 sources and NL AGN with 137) at $\overline{z}\sim 0.6$.
At the same redshift we found that BL
and NL AGN have a bias factor $\overline{b}_{BL}=1.62\pm 0.26$ 
and $\overline{b}_{NL}=1.56\pm 0.15$, which correspond to 
average halo masses $logM_{0}=13.27\pm 0.10[h^{-1}
M_{\odot}]$ and $12.97\pm 0.07[h^{-1}
M_{\odot}]$, respectively.\\
Similar results have been obtained using X-unobs and X-obs AGN samples;
unobscured AGN at $\overline{z}=1.16$ inhabit halos with average mass
$logM_{0}=13.30\pm 0.10 [h^{-1} M_{\odot}]$ which is higher at 
2.5 $\sigma$ level than the halo mass hosting obscured AGN 
($logM_{0}=12.97 \pm 0.08 [h^{-1} M_{\odot}]$), at similar redshift.\\
In order to compare our results with previous works on
the bias of X-ray selected AGN, we evaluated the bias factors corresponding
to the halo mass $M_{0}$ at $\overline{z}$ using \citet{She01} as shown
in Table \ref{tbl-2}, col  (5). 

Our results support the picture that at a given redshift, X-ray selected BL/X-unobs AGN
reside in more massive halos compared to X-ray selected NL/X-obs AGN.
This result would be expected if the two classes of AGN correspond to 
different phases of the AGN evolution sequence \citep{Hop06, Hop08, Hic09}.

\section{Bias Evolution and Constant Mass Threshold}
\label{sec:bevol}

\begin{figure*}
\plotone{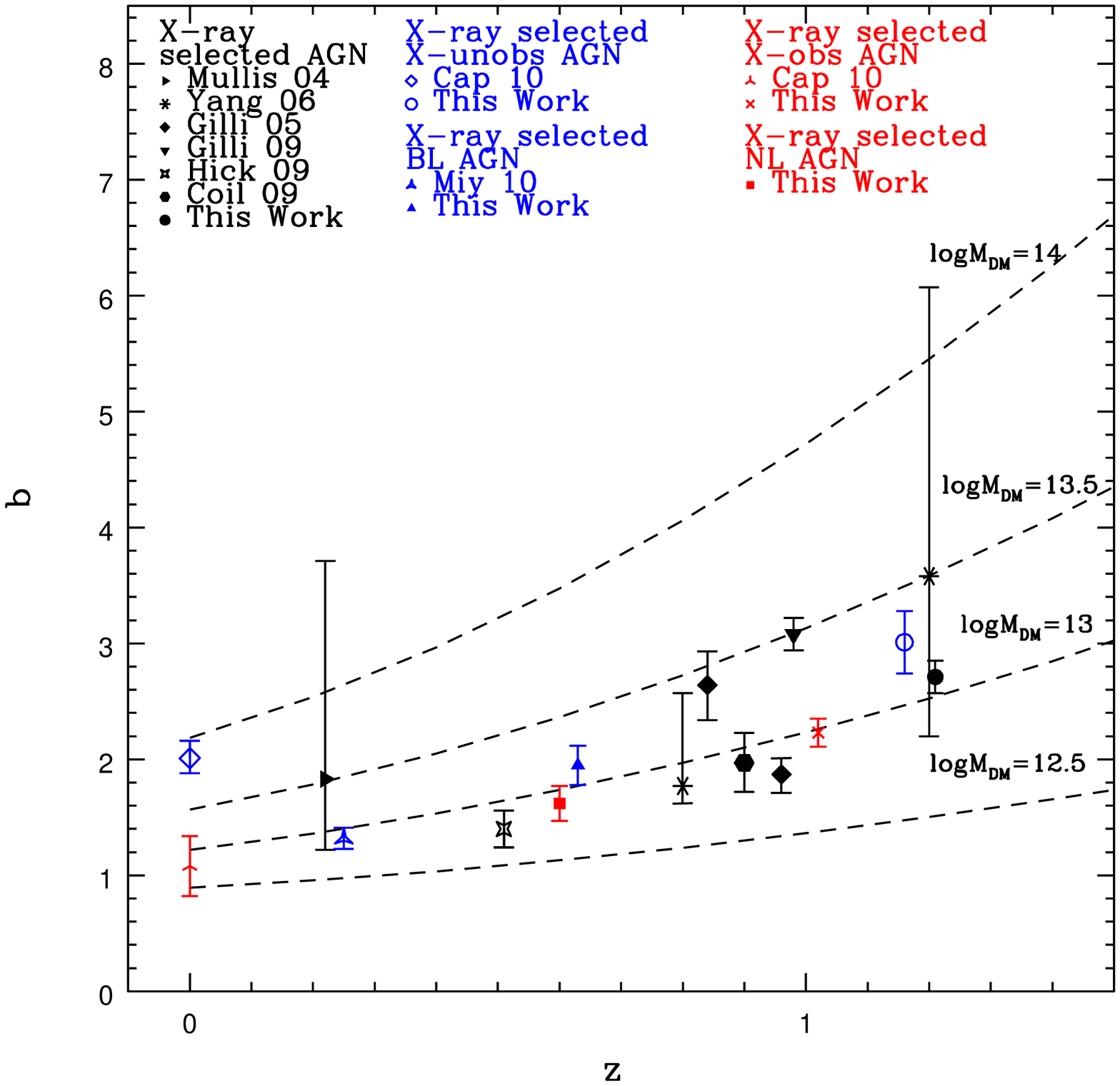}
\caption{\footnotesize Bias parameter as a function of redshift for various
  X-ray selected AGN (black data points), X-ray selected BL/X-unobs AGN
  (blue data points) and X-ray selected NL/X-obs AGN (red data points) as 
  estimated in previous studies and in this work according 
  to the legend. Our results refer to the bias factor $b_{S01}$ showed
  in Table \ref{tbl-2} col (5). The dashed lines show the expected
  $b(z)$ of typical DM halo masses $M_{DM}$ based on Sheth et al. (2001).
   The masses are given in $logM_{DM}$ in units of $h^{-1} M_{\odot}$. }
\label{fig:biasvsz}
\end{figure*}

\begin{figure*}
\plotone{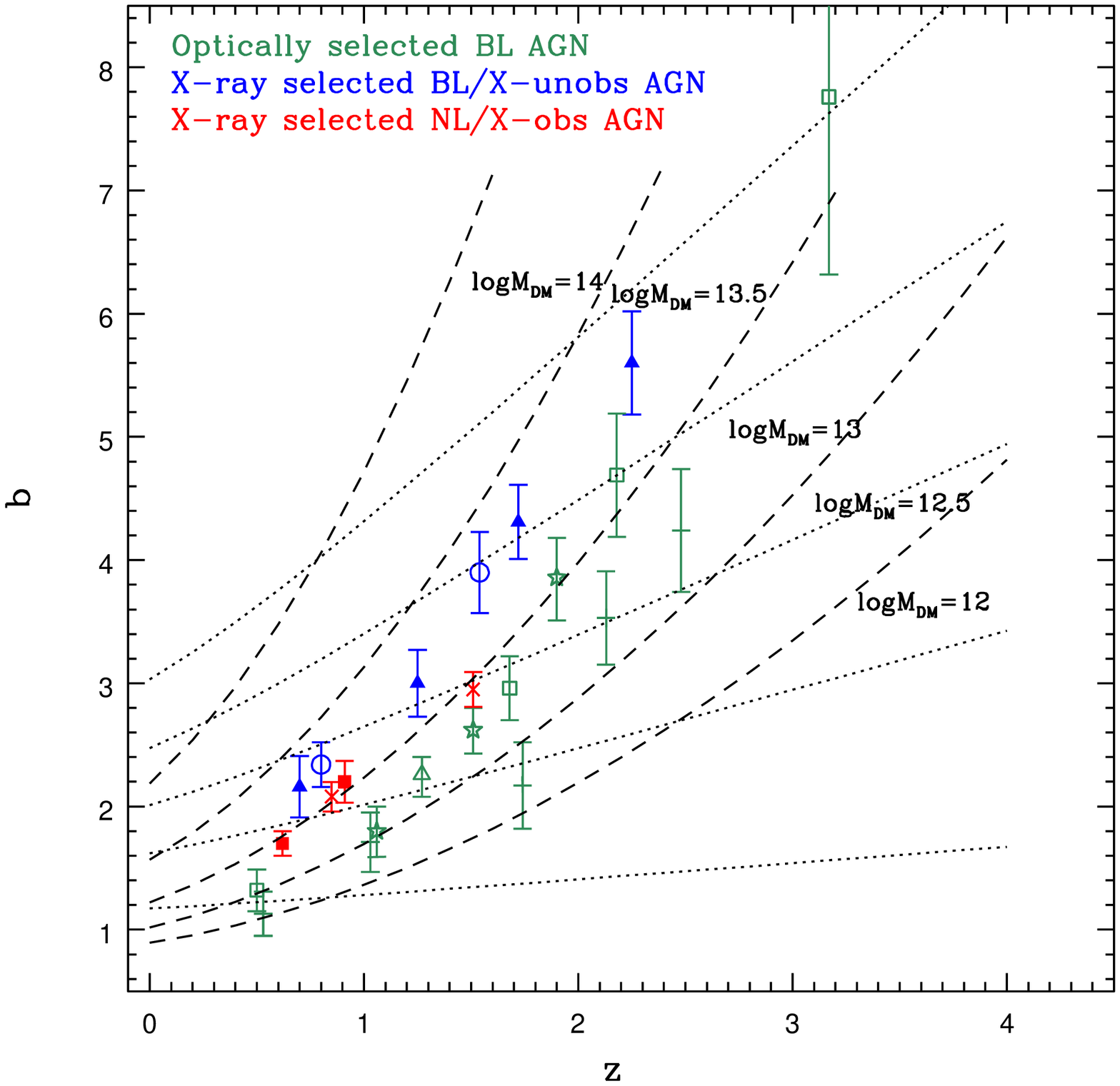}
\caption{\footnotesize Bias parameter as a function of redshift for
  optically selected BL AGN from previous works \citep{Cro05}, green-crosses;
  \citet{Por06}, green-stars; \citet{Shen09}, green-open squares; \citet{Ros09}, gree-open triangle) 
  and X-ray selected BL (blue triangles) and X-unobs (blue open-circles) AGN
  and NL (red squares) and X-obs (red crosses) AGN as estimated in this work. 
  The dashed lines show the expected $b(z)$ of typical DM
  halo masses based on \citet{She01} and the dotted
  lines represent the \emph{passive evolution} of the bias, as described in \citet{Fry96}. The bias of optically selected BL AGN 
  evolves with redshift following an evolution at constant halo mass, with a typical 
  mass which remains practically in the range $logM_{DM} \sim 12.5-13h^{-1}
  M_{\odot}$ at all redshifts $z<2.25$. X-ray selected BL/X-unobs AGN reside in more massive
  DM halos at all redshifts $z<2.25$, according to a typical mass of the hosting halos constant over time in
  the range $logM_{DM} \sim 13-13.5h^{-1} M_{\odot}$. The bias evolution
  of NL/X-obs AGN seems to indicate that they reside in DM halo mass $logM_{DM} \sim 13h^{-1}
  M_{\odot}$ constant at all $z<1.5$. These results suggest the 
  picture that X-ray selected BL AGN are triggered by secular processes as tidal disruption or disk instabilities instead of major mergers
  between gas-rich galaxies as confirmed by semi-analytic models and observations for optically selected quasars.}
\label{fig:biasvszall}
\end{figure*}
In order to investigate the redshift evolution of the bias factor, we split
the XMM-COSMOS AGN sample in three redshift bins. The sizes of the
redshift bins have been determined such that there are more or less 
the same number of objects in each bin. 
The values of $\overline{b}$, $\overline{z}$ and $M_0$
for the total AGN sample are shown in Table \ref{tbl-3}. 
The meaning of the table columns are: (1) sample; (2) number of sources; (3) bias parameter
from the projected DM 2-halo term, evaluated at the median $<z>$
of the sample; (4) typical halo mass using \citet{van02} and \citet{She01}; 
(5) weighted bias of the sample; 
(6) weighted redshift of the sample; (7) Average DM halo
mass; (8) Bias factor from $M_{0}$ estimated using Sheth et al. (2001).\\
We observed an increase of the AGN bias factor
with redshift, from $\overline{b}(\overline{z}=0.92)=1.80 \pm0.19$ to
$\overline{b}(\overline{z}=1.94)=2.63\pm0.21$ with a
DM halo mass consistent with being constant at $logM_{0}[h^{-1} M_{\odot}]\sim 13.1$
in each bin. These results support the picture that the bias of XMM-COSMOS AGN 
evolves with time according to a constant halo mass track at
all redshifts $z<2$.

This conclusion, based on the analysis of the global XMM-COSMOS AGN
sample, can however be affected by the fact that the relative proportions 
of BL and NL AGN are a strong function of redshift. In fact, since the 
XMM-COSMOS AGN sample is a flux limited sample, more luminous AGN 
are selected at high redshift and, also because of our magnitude 
limit, high-$z$ sources in our sample are mainly BL AGN (see \S\ref{fig:blnldistr}). 
For this reason BL AGN sample could be
analysed up to $z\sim2.25$, while the maximum average redshift of the two redshift 
bins for NL AGN is $z \sim 0.91$.
We found evidence of a strong increase of the BL AGN bias factor
in four redshift bins (see Table \ref{tbl-3}), with a DM halo 
mass constant at $logM_{0}[h^{-1} M_{\odot}]\sim13.28$ 
at all redshifts $z<2.25$.
For NL AGN we estimated $\overline{b}(\overline{z}=0.62)=1.59\pm0.13$ and
$\overline{b}(\overline{z}=0.91)=1.87\pm 0.19$, which correspond to
a constant halo mass values $logM_{0}[h^{-1} M_{\odot}] \sim 13.02$.
We split the X-unobs and X-obs AGN samples in two redshift bins up to
$\overline{z}\simeq 1.5$ and we found that the bias of X-unobs AGN (X-obs AGN) evolves
according to a constant halo mass consistent with the mass of BL AGN (NL AGN) hosting halos.
Fig. \ref{fig:bevol} (\emph{left panel}) shows the redshift evolution
of the average DM halo mass $M_{0}$ for all the AGN subsets. The horizontal lines
represent the mean value of $M_{0}$ for BL/X-unobs AGN (dashed-blue), 
NL/X-obs AGN (long dashed-red) 
and for the whole AGN sample (dotted-black).
Fig. \ref{fig:bevol} (\emph{right panel}) shows the redshift
evolution of the bias factors $b_{S01}$ (Table \ref{tbl-3}, col (7))
for different AGN sub-samples. 
The dashed lines show the expected 
$b(z)$ associated to the typical DM halo mass 
based on \citet{She01}.  \\
These results show that X-ray selected BL/X-unobs AGN reside in more massive
DM halos compared to X-ray selected NL/X-obs AGN at all redshifts $z$ at $\sim 3\sigma$ level.
This suggests that the AGN activity is a mass
triggered phenomenon and that different AGN phases
are associated with the DM halo mass, irrespective of redshift $z$.

\section{Discussion}
\label{sec:disc}

\subsection{Which DM halos host X-ray AGN?}

We have introduced a new method that uses the 2-halo
term in estimating the AGN bias factor and that 
properly accounts for the sample variance and the growth
of the structures over time associated with our use of large redshift interval
of the AGN sample.
Using this approach we have estimated an average 
mass of the XMM-COSMOS AGN hosting halos equal to
$logM_{0}[h^{-1}Mpc]=13.10 \pm 0.06$ which differs at 
$\sim 1.6 \sigma$ level from the typical halo mass $M_{DM}$
based on \citet{She01} using the methode 2 (see \S\ref{sec:2halo}).
The difference between the standard method and our own method
is also clear for the mass of BL and NL AGN hosting halos. 
We have found that BL AGN inhabit
DM halos with average mass $logM_{0}[h^{-1}Mpc]=13.24 \pm 0.06$ at 
$\overline{z}=1.53$ while halos hosting NL AGN have average mass
$logM_{0}[h^{-1}Mpc]=13.01 \pm 0.08$. 
BL AGN reside in more massive halos than NL AGN
also selecting two subsamples that
peak at the same median redshift $\overline{z} \sim 0.6$.
We obtained similar results using X-ray unobscured AGN at $\overline{z}=1.16$ and 
X-ray obscured AGN at $\overline{z}=1.02$ ($logM_{0}[h^{-1}Mpc]=13.30 \pm 0.10$
and $logM_{0}[h^{-1}Mpc]=12.97 \pm 0.08$, respectively).\\
Instead the typical halo mass
based on \citet{She01} using the AGN bias estimated
with the method 2, strongly depends on the median
redshift of the sample. According to the method 2, 
BL AGN at $<z>=1.55$ reside in less massive
halos compared to NL AGN at $<z>=0.74$, while the result is 
different selecting two samples of BL and NL AGN at the same $<z> \sim 0.5$.  
Our results agrees with the majority of the recent studies of X-ray surveys which
suggest a picture in which X-ray AGN are typically hosted in DM halos with mass in
the range $12.5<logM_{DM}[h^{-1}Mpc]<13.5$, at low ($<0.4$) and high ($\sim
1$) redshift. 
\citet{Sta10} found that \emph{Chandra/Bootes} AGN are located at 
the center of DM halos with $M>M_{min}=4 \times 10^{12}$ h$^{-1}$ $M_{\odot}$.
This mass estimate represents a threshold value, since they are assuming a halo
occupation described by a step function (zero AGN per halo/subhalo
below $M_{min}$ and one above it). Our approach, in terms of HOD, is
completely different. We assume a halo occupation described by $\delta$-function,
supported by the fact that AGN only reside in massive halos (then the AGN HOD can
be described by a narrow halo mass distribution at high mass values, but not by a step function).\\
Fig. \ref{fig:biasvsz} shows the bias factors
of X-ray selected AGN (black), BL/X-unobs AGN (blue) and NL/X-obs AGN (red)
as estimated in different surveys (according
to the legend). Our results refer to the bias factors $b_{S01}$ showed
in Table \ref{tbl-2}, column (5). The dashed lines 
show the expected $b(z)$ assuming a constant typical DM halo mass
$M_{DM}$, based on \citet{She01}. 

The previous studies of \citet{Gil05}
for the CDFN, \citet{Gil09}, \citet{Mul04}, \citet{Yan06}
for CLASXS AGN suggest the scenario in which the typical DM halo mass 
hosting X-ray selected AGN is $logM_{DM}[h^{-1} M_{\odot}] \sim 13.5$. 
The bias values measured in \citet{Gil05} on CDFS, in \citet{Hic09}, \citet{Coi09}
and \citet{Yan06} and in this work, correspond to
a lower halo mass ($logM_{DM}[h^{-1} M_{\odot}] \sim 13$).
A possible explanation could be that at fixed redshift, the bias
and then the mass of the hosting halo, depends on the luminosity of the
sample. The same explanation might be applied to the results on BL/X-unobs AGN.\\
The bias estimates at $z<1$ for NL/X-obs AGN in \citet{Cap10} and
in this work, seem to indicate that the mass of NL/X-obs AGN hosting halos is
$logM_{DM}[h^{-1}M_{\odot}] \sim 13$.

\subsection{Optically selected vs X-ray selected AGN}
\label{subsec:massthre}

We first found evidence of a redshift evolution of the bias factor
of X-ray selected BL/ X-unobs AGN (fig. \ref{fig:biasvszall}, blue data points)
and NL/X-obs AGN (red data points).
The bias evolves with redshift at constant average halo mass $logM_{0}[h^{-1} M_{\odot}] \sim 13.3$ for 
BL/X-unobs AGN  and $logM_{0}[h^{-1} M_{\odot}] \sim 13$ for NL/X-obs AGN 
at $z<2.25$ and $z<1.5$, respectively.
Fig. \ref{fig:biasvszall} shows the expected $b(z)$ assuming a constant typical DM
halo mass based on \citet{She01} (dashed lines)
and the so called \emph{passive bias evolution} \citep[dotted lines][]{Fry96}.
The observed bias evolution suggests an average halo mass of the hosting halos,
constant over time in the range $logM_{DM}[h^{-1} M_{\odot}] = 13-13.5$,
instead of an evolution of the bias in a model in
which objects are formed at a fixed time and their distribution evolves
under the influence of gravity.  

There have been several studies of the bias evolution of optical quasar
with the redshift as shown in fig. \ref{fig:biasvszall} (green data points), 
based on large survey samples such as 2QZ
and SDSS \citep{Cro05, Por06, Shen09, Ros09}. Since the quasar samples used in these clustering
analysis are defined as spectroscopically identified 
quasars with at least one broad (FWHM$>$1000 km $s^{-1}$) emission line, we refers
to them as optically selected BL AGN. 

All the previous studies infer the picture that the quasar
bias evolves with redshift following a constant mass evolution, with the average mass
that can vary in the range $logM_{DM}[h^{-1} M_{\odot}] \sim 12.5-13$, may be depending on the
AGN sample luminosity as already suggested for X-ray selected AGN.
The simplest interpretation according to the observed
redshift evolution of the bias factors is that 1) X-ray selected AGN whether 
BL/X-unobs or NL/X-obs AGN inhabit DM halos with mass
higher than the mass of optically selected quasar hosting halos in the range $z=0.5-2.25$;
2) X-ray selected BL/X-unobs AGN  
reside in more massive halos compared to NL/X-obs AGN for $z=0.6-1.6$
and the discrepancy between the bias factors of the two samples increases with $z$;
3) the AGN activity is a mass triggered 
phenomena and the different AGN 
evolutionary phases are associated with just the DM halo mass, irrespective of the
redshift $z$.

\subsection{External vs Internal Triggering}

\begin{figure}
\plotone{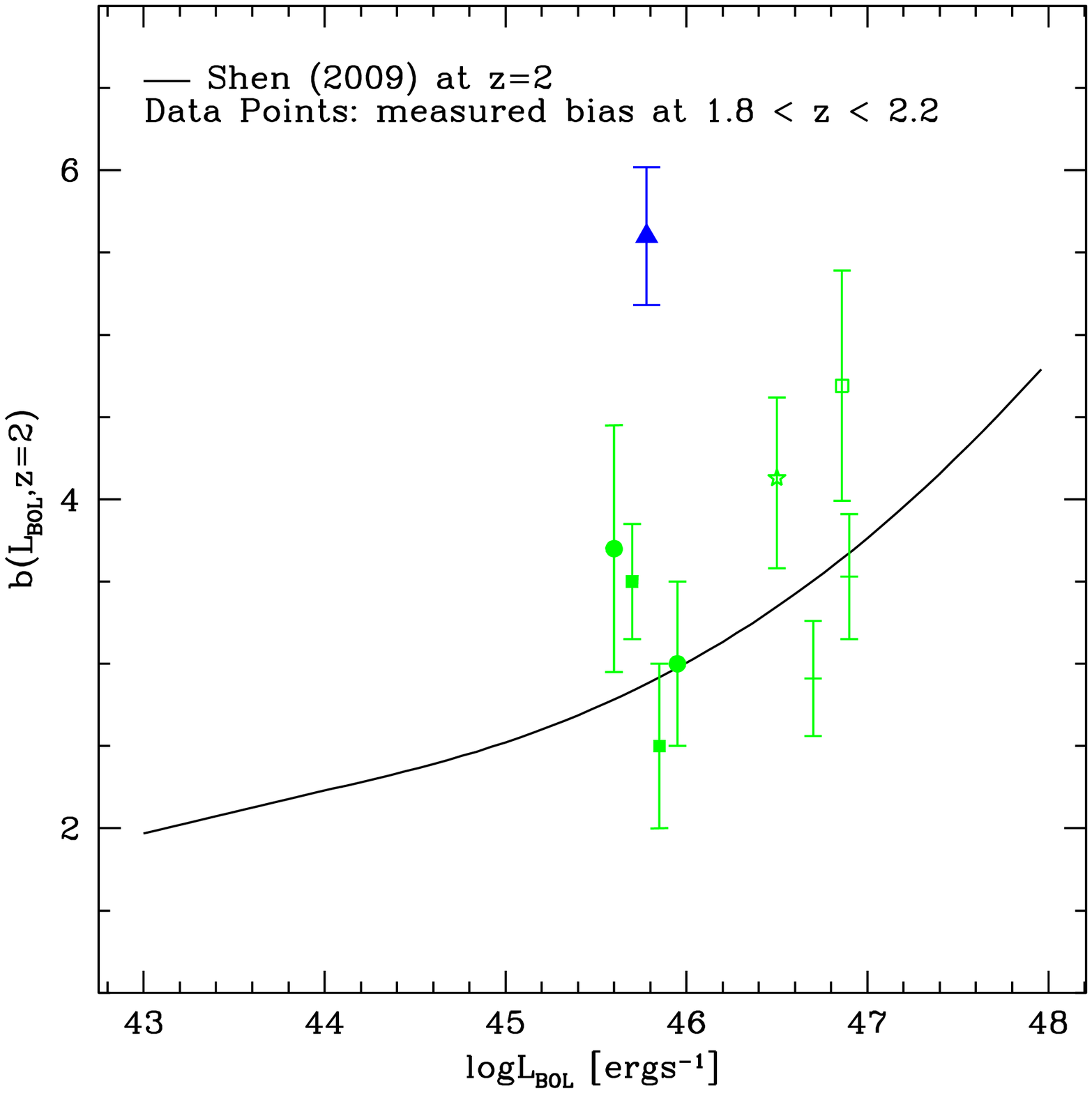}
 \caption{\footnotesize Predicted bias as a function of luminosity, computed according
 to \citet{Shen09a} fixing $z=2$, compared to previous bias estimates at $1.8<z<2.2$,
 for optically selected BL AGN and for
 XMM-COSMOS BL AGN. Points are measurements from \citet[green-crosses]{Cro05},
 \citet[green-star]{Por06}, \citet[green-open square]{Shen09}, \citet[green-circles]{Ang08},
 \citet[green-squares]{Mye07} and our result (blue triangle). For ease of comparison,
 all luminosities are converted to bolometric luminosities using the
 corrections from \citet{Hop07}.
 The theoretical model which assumes a quasar phase triggered by
 major merger reproduces the results obtained 
 for the bias of quasars, but can not reproduce the high
 bias factors found for X-ray selected BL AGN and then can not
 explain why optically selected quasars
 that have higher bolometric luminosity compared to COSMOS X-ray selected BL/X-unobs AGN,
 reside in more massive halos.
 These differences suggest a switch to a different dominant mechanism
 for AGN triggering, from major mergers between gas-rich galaxies to 
 secular processes as tidal disruptions or disk instabilities.}
\label{fig:biasShen}
\end{figure}

The major merger of galaxies is one of the promising mechanisms suggested 
to be responsible for fuelling quasars and in particular to be dominant for
bright quasars at high redshift. Models of major mergers appear
to naturally produce many observed properties
of quasars, as the quasar luminosity density, the shape and the evolution
of the quasar luminosity function and the large-scale quasar clustering as
a function of $L$ and $z$ \citep{Hop08, Shen09a, Sha09, Sha10, Sha10rev, Bon09}. \\
Clear evidence for higher incidence of mergers is seen among quasars \citep{Ser06, Hop06, Vei09}.
Additionally a large fraction of luminous quasars at low redshift are 
associated with either morphologically disturbed objects \citep{Can01, Guy06},
or early-type hosts with fine structure in their optical light
distribution, indicative of past interactions \citep{Can07, Ben08}. In the local
Universe, for instance, the study of the environment of Swift BAT Seyfert galaxies 
\citep{Kos10} appeared to show an apparent mergers 
$\sim 25\%$ which suggests that AGN activity and 
merging are critically linked.
Moreover it is believed that major merger dominates at high
redshift and bright luminosities \citep{Has08, Hop06}, while
minor interaction or bar instabilities or minor
tidal disruptions are important at low redshift ($z\lesssim 1$) and low luminosities
($L_{BOL}\lesssim 10^{44}erg$ s$^{-1}$) \citep{Hop09}.

Our results on the bias evolution of X-ray selected BL/X-unobs AGN
infer that these objects with $L_{BOL}\sim 2 \times 10^{45}erg$ $s^{-1}$ reside in massive DM halos
$M_{DM} \sim 2 \times 10^{13}M_{\odot}h^{-1}$. 
Besides studies on BL AGN in the COSMOS field \citep{Mer10,Tru11}
suggest that our sample is characterized by BH masses 
in the range $M_{BH}=10^7-10^{9}M_{\odot}$ 
and Eddington ratio $\lambda>0.01$.
Optically selected quasars from large survey samples such as 2QZ
and SDSS are high-luminosity quasars $L_{BOL}\gtrsim 10^{46}erg^{-1}$
with BH masses in the range $M_{BH}=10^8-10^{10}M_{\odot}$ and $\lambda>0.01$.
Clustering analysis of optical quasars 
have shown that they reside in DM halos with $M_{DM}\sim10^{12}M_{\odot}$ $h^{-1}$.

Fig. \ref{fig:biasShen} shows the predicted bias as a function of luminosity computed
according to \citet{Shen09a} at $z=2$. The theoretical model which assumes a quasar
phase triggered by major mergers predicts an increasing bias with luminosity
and reproduces the previous results obtained 
for optical quasars at $1.8<z<2.2$ (\citet[green-crosses]{Cro05}, \citet[green-star]{Por06}, 
\citet[green-open square]{Shen09}, \citet[green-circles]{Ang08}, \citet[green-squares]{Mye07}).
On the other hand the model can not reproduce the high
bias factor found for X-ray selected COSMOS BL AGN (blue triangle) and then can not
explain why optically selected quasars
characterized by higher bolometric luminosity compared to 
X-ray selected COSMOS BL/X-unobs AGN,
are found in less massive halos. 
These differences suggest a switch to a different dominant mechanism
for AGN triggering. 

\citet{HoHe06} introduced a model for the
fueling of low-luminosity AGN (Seyferts, with $L_{BOL}\lesssim 10^{44}-10^{45}erg$ $s^{-1}$ and 
$M_{BH} \lesssim 10^7 M_{\odot}$), which proposes
AGN triggered by random accretion of gas via internal, secular processes.
The stochastic accretion model and the merger-driven activity
are fundamentally different, the former being determined by stochastic
encounters with a cold gas supply in a quiescent system, 
the latter by the violent torquing of cold gas throughout entire galaxies into 
the galaxy center in major mergers.
Accretion of cold gas in quiescent systems can account for low luminosity Seyferts 
but can not explain the higher luminosities and the larger BH masses observed 
for XMM-COSMOS BL AGN. The high Eddington ratios
at masses in the range $M_{BH} \sim 10^8-10^9 M_{\odot}$ can not be maintained
through this mode of accretion.\\
Furthermore, this fueling mechanism predicts lower bias factors compared to the
major merger picture for bright quasars, 
which is completely in disagreement with our results.

Fueling by stellar winds or hot gas accretion may represent
yet a third qualitatively distinct mode of fueling.
\citet{Cio97,Cio01} investigated the episodic AGN activity
model in early-type galaxies, assuming at their center
the presence of a massive BH growing with the accretion
of matter and affecting the inflow through feedback.
The duration of the single accretion event are extremely short 
but the maximum luminosities reached during the accretion events can be of the
order of $L_{BOL}\sim 10^{46}-10^{47}erg$ $s^{-1}$,
depending on the input parameters of the model.
The central BH grows by episodic accretion up to a mass
in the observed range ($M \sim 10^{8.5}-10^{9.5}M_{\odot}$)
in all giant ellipticals.\\
On the other hand the observational consequence of this model is that
the duty cycle is very low, typically of the order of $10^{-2}-10^{-3}$.
This result implies a small fraction of giant ellipticals observed
in an AGN phase, too low compared to the
observed 10\% of X-ray AGN residing in massive galaxies.

In the AGN evolutionary model described in Hickox et al. (2009),
optically bright quasars are hosted by ongoing disk galaxy
mergers and immediately precede an optically faint X-ray AGN phase, which
evolves into an early-type galaxy. Following this evolutionary sequence, NL/X-obs
AGN should be triggered in the first initial phase of vigorous star
formation and obscured accretion which supports the scheme of NL AGN inhabiting
halos with low typical masses $logM_{DM}[h^{-1} M_{\odot}] \sim
12.5$.  An X-ray AGN phase immediately follows the quasar phase. Since DM halos grow
and accumulate mass over time, X-ray AGN reside in more massive DM halos with
typical mass $logM_{DM}[h^{-1} M_{\odot}] \sim 13-13.5$. 
This model predicts that X-ray AGN reside in more massive halos than QSO,
but assumes a decline of the BH accretion rate from its peak in the quasar phase to
$\dot{M}\lesssim 10^{-2}\dot{M_{Edd}}$ or lower, which is in disagreement with \textit{the high 
Eddington ratios} found for XMM-COSMOS BL AGN \citep{Mer10, Tru11}.

A plausible scenario requires 
that high-luminosity quasars ($L_{BOL} > 10^{46}erg$ $s^{-1}$)
are triggered by external processes such as major mergers between gas-rich galaxies with masses 
of the order of $M_{\ast} \sim 10^{10}M_{\odot}$.
Instead for BL AGN with $L_{BOL}\sim 2 \times 10^{45}erg$ $s^{-1}$,
internal mechanisms such as tidal disruptions or disk instabilities
in massive galaxies ($M_{\ast} \sim 10^{11}M_{\odot}$) 
might play a dominant role.

The morphology of the AGN hosts galaxies provides an important clue 
into the mechanism that triggers their current AGN activity.
It was observed that many AGN are not fueled by major mergers and only
a small fraction of AGN are associated with morphologically disturbed
galaxies. 
\citet{Cis10} analysed a sample of X-ray selected AGN host galaxies
and a matched control sample of inactive galaxies in the COSMOS field.
They found that mergers and interactions involving AGN hosts are not dominant 
and occur no more frequently than for inactive galaxies. Over 55\% of the studied
AGN sample which is characterized by $L_{BOL} \sim 10^{45} erg$ $s^{-1}$ and by 
mass of the host galaxies $M_{\ast} \gtrsim 10^{10}M_{\odot}$
are hosted by disk-dominated galaxies.
This high disk fraction means that the lack of disturbed morphologies observed 
among the AGN hosts can not simply be due to a time lag between merger 
activity and X-ray visibility and suggests that secular fueling 
mechanisms can be high efficient.\\
It was also suggested by \citet{Geo09} that bar
instabilities and minor interactions are more efficient in producing
luminous AGN at $z\lesssim 1$ and not only Seyfert galaxies and
low-luminosity AGN as the \citet{HoHe06,Hop09} model predicts.
Besides several works on the AGN host galaxies \citep{Dun03, Gro05,
Pie07, Gab09, Rei09, Tal09} show that the morphologies of the AGN 
host galaxies do not present a preference for merging systems. 

At the redshift of our interest, recent findings of \citet{Sch11} and 
\citet{Ros11}, who examined a smaller sample of AGN at $z\sim 2$ 
in the ERS-II region of the GOODS-South field,
inferred that late-type morphologies are prevalent among the AGN hosts.
The role that major galaxy mergers play in triggering AGN activity at 1.5 $< z <$ 2.5
was also studied in the CDF-S. Kocevski et al. (in prep.) found that
X-ray selected AGN at $z\sim 2$ do not exhibit a significant excess 
of distorted morphologies while a large fraction reside in late-type galaxies.
They also suggest that these late-type galaxies are fueled by the stochastic 
accretion of cold gas, possibly triggered by a disk instability or minor interaction.

We want to stress that our results by no means infer that mergers 
make no role in the AGN triggering. On the contrary, high luminosity AGN and 
probably a fraction of moderate luminosity AGN
in our sample might be fuelled by mergers. In fact, given the complexity of AGN
triggering, a proper selection of an AGN sub-sample, using for instance the luminosity,
can help to test a particular model boosting the fraction of AGN host 
galaxies associated with morphologically disturbed galaxies. 

Our work might extend the statement that for moderate luminosity X-ray selected BL AGN secular processes 
might play a much larger role than major mergers up to $z \sim 2.2$, compared to the previous 
$z \lesssim 1$, even during the epoch of peak merger-driven accretion.

\section{Conclusions}
\label{sec:conc} 

We have studied the redshift evolution of the bias factor of 593 XMM-COSMOS AGN with
spectroscopic redshifts $z<4$, extracted from the 0.5-2 keV
X-ray image of the 2$deg^2$ XMM-COSMOS field. We have described a new 
method to estimate the bias factor and the associated DM halo
mass, which accounts for the growth of the structures
over time and the sample variance. Key results can be summarized as follows:
\begin{enumerate}
\item We estimated the AGN bias factor  
  $b_{S01}=2.71 \pm 0.14$ at $\overline{z}=1.21$ which 
  corresponds to a mass of DM halos hosting AGN 
  equal to $logM_{0}[h^{-1}
  M_{\odot}] = 13.10\pm 0.10$. 
 \item We split the AGN sample in broad optical emission lines AGN (BL) and AGN without
  optical broad emission lines (NL) and for each of them we considered a subset with 
  $\overline{z}=0.6$ and we found that
  BL and NL AGN present $b_{S01}=1.95\pm
  0.17$ and $b_{S01}=1.62\pm 0.15$, which correspond to masses equal to
  $logM_{0}[h^{-1} M_{\odot}]=13.27\pm 0.10$ and
  $12.97\pm0.07$, respectively. 
 \item We selected in the hard band a sample of X-ray unobscured and X-ray obscured AGN
 according to the column density and we found that X-ray unobscured (X-ray obscured) AGN inhabit 
 DM halos with the same mass compared to BL (NL) AGN with 
 $logM_{0}[h^{-1} M_{\odot}]=13.30\pm 0.10$ ($logM_{0}[h^{-1} M_{\odot}]=12.97\pm 0.08$).
\item We found evidence of a redshift evolution of the 
bias factors for the different AGN subsets, corresponding to a constant DM halo
mass threshold which differs for each sample. XMM-COSMOS AGN are
hosted by DM halos with mass $logM_{0}=13.12 \pm 0.07 [h^{-1} M_{\odot}]$
constant at all $z<2$, BL/X-ray unobscured AGN reside in halos with mass 
$logM_{0}=13.28 \pm 0.07 [h^{-1} M_{\odot}]$ for $\overline{z}<2.25$
while XMM-COSMOS NL/X-ray obscured AGN inhabit less massive halos  
$logM_{0}=13.00 \pm 0.06 [h^{-1} M_{\odot}]$, constant at all $\overline{z}<1.5$.
\item The observed bias evolution for XMM-COSMOS BL and NL AGN at all $z<2.25$,
suggests that the AGN activity is a mass triggered phenomenon and that different AGN 
evolutionary phases are associated with just the DM halo mass, irrespective of the
redshift $z$. 
\item The bias evolution of X-ray selected BL/X-ray unobscured AGN corresponds to halo 
masses in the range $logM_{DM}[h^{-1} M_{\odot}] \sim 13-13.5$ 
typical of poor galaxy groups at all redshifts.
Optically selected BL AGN instead reside in lower density environment with 
constant halo masses in the range $logM_{DM}[h^{-1} M_{\odot}] \sim 12.5-13$
at all redshifts. This indicates that X-ray and optically selected AGN do not 
inhabit the same DM halos.
\item The theoretical models which assume a quasar phase triggered by
 major mergers can not reproduce the high
 bias factors and DM halo masses found for X-ray selected BL AGN up to $z\sim 2.2$.
 Our results might suggest the statement that for moderate luminosity X-ray selected BL AGN 
 secular processes such as tidal disruptions or disk instabilities 
 play a much larger role than major mergers up to $z \sim 2.2$, compared to the previous 
 $z \lesssim 1$.
 \end{enumerate}

\acknowledgments

VA, GH \& MS acknowledge support by the German Deutsche Forschungsgemeinschaft,
DFG Leibniz Prize (FKZ HA 1850/28-1). FS acknowledges support from the 
Alexander von Humboldt Foundation.

\clearpage

\end{document}